\documentclass{WileyMSP-template}
\usepackage[normalem]{ulem}
\usepackage{mathrsfs}
\usepackage{mhchem}
\usepackage{makecell}
\usepackage{color,xcolor,verbatim}

\usepackage{multirow}
\usepackage{booktabs}
\newcommand{\clr}{\color{red}}

\newcommand{\bee}{\begin{equation}}
\newcommand{\ee}{\end{equation}}
\newcommand{\bma}{\begin{pmatrix}}
\newcommand{\ema}{\end{pmatrix}}
\newcommand{\balig}{\begin{align}}
\newcommand{\ealig}{\end{align}}

\begin{document}

\pagestyle{fancy}
\rhead{\includegraphics[width=2.5cm]{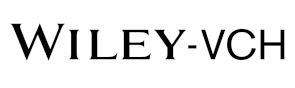}}

\title{Catalog of  Altermagnetism in Magnetic Wallpaper/Space Groups and Nonsymmorphic Altermagnets}

\maketitle


\author{Congcong Le}
\author{Fan Cui}
\author{Iao-Fai Io}
\author{Moritz M. Hirschmann}
\author{Xianxin Wu}
\author{Ching-Kai Chiu}



\begin{affiliations}

Congcong Le\\
Hefei National Laboratory, Hefei 230088, China\\
School of Emerging Technology, University of Science and Technology of China, Hefei 230026, China\\
Email Address:lecongcong@hfnl.cn

Congcong Le, Iao-Fai Io, Ching-Kai Chiu\\
RIKEN Center for Interdisciplinary Theoretical and
Mathematical Sciences (iTHEMS), RIKEN, Wako 351-0198, Japan\\
Email Address:ching-kai.chiu@riken.jp

Iao-Fai Io\\
Department of Physics and Center for Theoretical Physics, National Taiwan University, Taipei 10607, Taiwan\\

Moritz Hirschmann\\
RIKEN Center for Emergent Matter Science (CEMS), Wako, Saitama 351-0198, Japan \\

Fan Cui, Xianxin Wu\\
CAS Key Laboratory of Theoretical Physics, Institute of Theoretical Physics, Chinese Academy of Sciences, Beijing 100190, China\\
Email Address:xxwu@itp.ac.cn

\end{affiliations}



\begin{abstract}

Conventional altermagnetism, characterized by compensated collinear spin alignment and spin splitting, exhibits identical spin states at opposite momenta. In this work, we employ a non-spatial global symmetry $S$, the spinless time-reversal symmetry, which effectively replaces inversion symmetry in preserving the spin-state equivalence; hence, we systematically extend the classification of altermagnetism to all possible non-centrosymmetric crystals. By analyzing the necessary symmetry conditions, we provide a complete catalog of altermagnetic orders for all 2D magnetic wallpaper groups and all 3D magnetic space groups, identifying 17 altermagnetic wallpaper groups (12 centrosymmetric and 5 non-centrosymmetric) and 422 altermagnetic space groups (160 centrosymmetric and 262 non-centrosymmetric). This catalog assigns each altermagnetic wallpaper and space group to one of the six altermagnetic wave types established in the literature and presents its distinct spin distribution in the Brillouin zone (BZ); notably, the low-energy wave-type description does not necessarily extend throughout the full BZ, since the spin-degenerate nodal lines and planes can be unpinned from the high-symmetry planes. Beyond the catalog, nonsymmorphic symmetries further bring new patterns of the altermagnetic BZs through the emergence of hourglass dispersions, which arise from the compatibility relations between two symmetry-protected degenerate manifolds: same-spin and opposite-spin degeneracies. In both the non-centrosymmetric altermagnetism and the emergence of the hourglass dispersion, the spinless time-reversal symmetry plays the key role. Our work extends the symmetry catalog of altermagnetism and reveals that nonsymmorphic symmetries are essential for realizing altermagnetic band structures beyond the six established wave types, such as an $i$-wave-like spin winding in a tetragonal BZ.

\end{abstract}


\section{Introduction}
Altermagnetism has been recognized as a third fundamental collinear magnetic phase, distinct from traditional ferromagnetism and antiferromagnetism, and has attracted enormous attention in condensed matter physics\cite{Hayami2019,Libor2022a,Libor2022b,krempasky2024,cheong_altermagnetism_2025}. Unlike ferromagnets, which exhibit finite net magnetization, and antiferromagnets, where time-reversal symmetry combined with inversion symmetry or lattice translations preserves spin degeneracy,  altermagnets host a compensated magnetic order in which all spins are aligned along a single common axis. Despite the zero net magnetization, the spin degeneracy is lifted away from the time-reversal-invariant momenta. Remarkably, this combination of a compensated magnetic order and lifted spin degeneracy gives rise to nonrelativistic momentum-dependent spin-splitting in the Brillouin zone~(BZ) in the absence of spin-orbit coupling (SOC).
The spin-splitting electronic structure makes altermagnets a distinct class of materials with fascinating properties\cite{shao2021,Gonz2021,Tomas2022,bose2022,smejkal2022,feng2022,bai2022,Karube2022,Gonzalez2023,Bai2023,zhou2024,parshukov2024}, such as anomalous Hall effect, nonrelativistic spin current, thermal transport, spin-splitting torque phenomena, and topological band responses. To capture the essential properties of altermagnetism, minimal models have been developed for all centrosymmetric space groups \cite{Roig2024}, providing a theoretical framework for its characterization. Moreover, the interplay between altermagnetism and superconductivity has been proposed to give rise to several intriguing phenomena, including topological superconductivity\cite{Zhu2023,Brekke2023,Xuan2023,Ghorashi2024}, Andreev reflection\cite{Papaj2023}, diode effects\cite{Banerjee2024}, finite-momentum Cooper pairing\cite{Debmalya2024,zhang2024} and Josephson effects\cite{Ouassou2023,Beenakker2023}. Beyond collinear systems, the concept of altermagnetism has also been extended to non-collinear altermagnets~\cite{Hayami2020,cheong_altermagnetism_2024,zhu_observation_2024}, further broadening its conceptual scope and potential material realizations. Recently, an odd-parity $p$-wave magnetic counterpart, which lifts Kramers degeneracy without requiring collinear order and necessarily violates the identical-spin-at-opposite-momenta condition, has also been proposed~\cite{Hellenes2023} and experimentally realized\cite{Song2025, Yamada2025}; throughout this work, however, we restrict ourselves to the conventional collinear definition of altermagnetism in the absence of SOC.

\begin{figure}
\centerline{\includegraphics[width=0.5\textwidth]{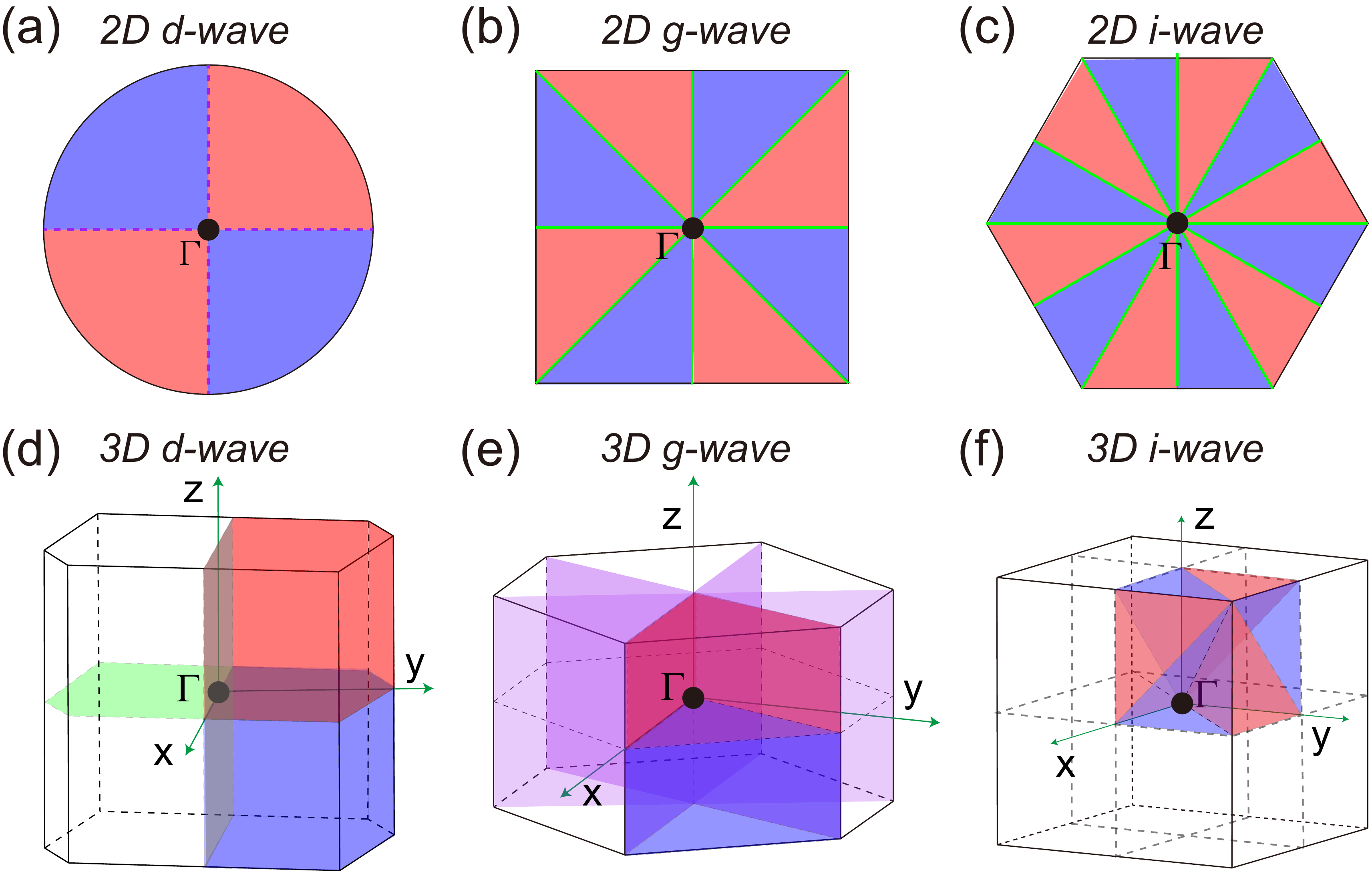}}
\caption{(color online) Classification of 2D and 3D altermagnetic systems, with and without inversion symmetry, into three fundamental types:  $d$-wave, $g$-wave and $i$-wave. The red and blue regions represent the spin-up and spin-down distributions in BZs, and the conventional-cell BZs are used. (a) A circular representation is adopted for 2D $d$-wave, indicating that its BZ does not have a fixed shape. It can appear in the BZs of 2D and 3D orthorhombic and square lattices. (b) 2D $g$-wave is confined to square lattices in 2D and 3D systems. (c) 2D $i$-wave appears only in trigonal and hexagonal lattices in 2D and 3D systems. (d) 3D $d$-wave is only in monoclinic lattice. (e) 3D $g$-wave is restricted to trigonal and hexagonal lattices. (f) 3D $i$-wave occurs exclusively in cubic lattice. The blue solid lines and black dashed lines in panels (a)-(c) indicate spin-degenerated nodal lines. The blue solid lines are protected by mirror symmetry and thus fixed in momentum space. In contrast, the black dashed lines, which may be unprotected in square lattices, can shift or deform, while in orthorhombic lattices they remain symmetry-protected, indicating that whether these nodal lines are fixed depends on the lattice symmetries. The green, gray, and purple planes in panels (d)-(f) represent spin-degenerate nodal planes. The green planes are symmetry-protected and thus fixed; the gray planes are unprotected and can deform; whether the purple planes are fixed depends on the crystal symmetries.
\label{types} }
\end{figure}


The basic classification of altermagnetism has been laid out at the magnetic-point-group level\cite{Libor2022a,Libor2022b}, where momentum-dependent spin splittings are organized into six fundamental wave types --- 2D $d$-, $g$-, $i$-wave and 3D $d$-, $g$-, $i$-wave (Fig.~\ref{types}) --- and the corresponding magnetic point groups, predominantly centrosymmetric and more briefly non-centrosymmetric, have been enumerated. Subsequent group-theoretical studies have extended this framework through Landau theories of altermagnetic order~\cite{McClarty2024,Schiff2025}, a tensorial classification of altermagnetic point groups~\cite{Radaelli2024}, spin-layer-group categorizations of 2D altermagnetism~\cite{Zeng2024,Zeng2026FoP}, and a unified symmetry classification of magnetic orders based on oriented spin space groups~\cite{Liu2026}. For visual clarity in Fig.~\ref{types}(d)-(f), only a partial region of the BZ is depicted; the full spin-splitting structure follows by applying the relevant rotational and mirror symmetries. Guided by this framework, a wide range of centrosymmetric altermagnetic materials have been proposed and observed\cite{Libor2022a,Libor2022b,Ahn2019,Yuan2020,Mazin2021,Lee2024,Reimers2024,yang_three-dimensional_2025,jiang2025}, including the rutile metals \ce{RuO2}~(2D $d$-wave), \ce{MnF2}~(2D $d$-wave), \ce{FeSb2}~(2D $d$-wave), \ce{MnTe}~(3D $g$-wave), \ce{CrSb}~(3D $g$-wave), and \ce{KV2Se2O}, and an AI-driven automated discovery approach\cite{gao2024} has further predicted 50 additional centrosymmetric candidates. In addition, a large-scale ab initio screening of the MAGNDATA database has identified about sixty spin-split collinear antiferromagnets, encompassing both centrosymmetric and non-centrosymmetric structures~\cite{Guo2023}. Similarly, a magnetic-space-group survey has tabulated the collinear antiferromagnets exhibiting momentum-dependent spin splitting, comprising 160 centrosymmetric and 262 non-centrosymmetric groups~\cite{Yuan2021PRM}. Building on this foundation, this manuscript makes two systematic refinements that go beyond the prior point-group treatment. First, we provide an exhaustive catalog of altermagnetic orders for all 2D magnetic wallpaper groups and all 3D magnetic space groups by analyzing the necessary symmetry conditions, encompassing both centrosymmetric and non-centrosymmetric cases. Moreover, we expose the role of nonsymmorphic symmetries that magnetic-point-group classifications cannot capture, which is a fundamentally new class of altermagnetic spin-splitting beyond the conventional 'six-wave' classification. Second, whereas Fig.~\ref{types} depicts the six wave types in the conventional-cell BZ, we further classify the spin-momentum textures in the primitive-cell BZ, in which two altermagnetic wallpaper (or space) groups belonging to the same altermagnetic point group can host qualitatively distinct spin-splitting patterns --- for example, the pair P4$^{\prime}$m$^{\prime}$m and P4$^{\prime}$mm$^{\prime}$ at the wallpaper-group level, whose spin-degenerate nodal lines lie along the $k_x/k_y$ axes and along the BZ diagonals, respectively --- yielding a substantially richer classification. Together, this wallpaper-/space-group-level catalog in the primitive-cell BZ both significantly expands the landscape of altermagnetic materials and provides deeper insight into the fundamental mechanisms governing altermagnetism.

 A central ingredient throughout this manuscript is a non-spatial global symmetry $S$.  Because $S$ does not exchange the two spin sectors but does reverse momentum, it acts within each block of the spin-diagonal Hamiltonian as an effective spinless time-reversal symmetry. Such an operation has been introduced in the literature as the construction underlying spin-group classifications of altermagnetism\cite{Libor2022a,Libor2022b}. Globally, in noncentrosymmetric altermagnets, $S$ therefore plays the role that inversion symmetry plays in centrosymmetric altermagnets, preserving identical spin states at opposite momenta even when crystal inversion is absent; consequently, the non-centrosymmetric altermagnetic systems we identify are symmetry-equivalent to their centrosymmetric counterparts and host the same six wave types of $d$-, $g$-, and $i$-wave splittings (Fig.~\ref{types}).

 The interplay between altermagnetism and nonsymmorphic symmetries has been touched upon in the recent literature: Roig \emph{et al.}\cite{Roig2024} construct minimal tight-binding models for centrosymmetric space groups and provide insight into why most altermagnets are realized in nonsymmorphic groups, while Fakhredine \emph{et al.}\cite{Fakhredine2023PRB108_115138} examine a single Pnma example and show that nonsymmorphic symmetries enforce a fourfold degeneracy on the BZ boundary which generates a large anomalous Hall effect through semi-Dirac anticrossings. Similarly, a magnetic-space-group and density-functional study of two nonsymmorphic Pnma perovskites (\ce{BiFeO3} and \ce{CaMnO3}) finds a conventional $d$-wave spin splitting~\cite{Rooj2025}. These works, however, remain within the conventional spin-splitting taxonomy.

 In contrast, this manuscript identifies a genuinely new class of altermagnetic spin-splitting BZ that emerges when $S$ is composed with a nonsymmorphic spatial operation. For a glide mirror or a screw axis $\tilde{g}$, the interplay between the same-spin degeneracy enforced by the composite antiunitary symmetry $\tilde{g}S$ and the opposite-spin degeneracy nodes from the conventional altermagnetism can lead to symmetry-enforced hourglass dispersions---an idea inherited from the hourglass-fermion topological insulators\cite{wang_hourglass_2016,Shiozaki2015PRB91_155120,Wieder2018Science361_246}---and the resulting band crossings appear as nodal lines in 2D wallpaper groups and as nodal surfaces in 3D space groups. Such hourglass-induced spin textures fall outside the conventional six-wave taxonomy and constitute a new class of altermagnetic spin-splitting BZ that is exhaustively classified in this work, with P4$^\prime$gm$^\prime$ and its 3D extension P4$^\prime$bm$^\prime$ and ${P4_2'2_12'}$ as prototypical examples.



\section{Altermagnetism Catalog Preliminary}

Before systematically exploring wallpaper and space groups for altermagnetism, we establish the selection principles based on the conventional definition of altermagnetism:
\begin{enumerate}
    \setcounter{enumi}{0}
    \item \textbf{Momentum-dependent spin splitting:} This spin non-degeneracy differs from antiferromagnetism.
    \item \textbf{Absence of spin mixing:} Negligible SOC.
    \item \textbf{Collinear magnetization:} Spins are aligned along a single axis (spin-up and spin-down).
    \begin{enumerate}
        \item Spin polarization must be present.
        \item Non-magnetic sites are permitted but not required.
    \end{enumerate}
    \item \textbf{Compensated magnetic order:} Zero net magnetization.
    \item \textbf{Identical spin at opposite momenta:} This applies only to the conventional cases.
\end{enumerate}
In the literature, Condition 2 corresponds to the non-relativistic limit and $p$-wave altermagnetism naturally violates Condition 5~\cite{Hellenes2023,Song2025,Yamada2025}, but this manuscript focuses on the conventional definition by keeping these five conditions. As a result of Conditions 2 and 3, the free-fermion Hamiltonian can be block-diagonalized into decoupled spin-up and spin-down sectors:
\bee
H=\bma
H_{\uparrow} & 0 \\
0 & H_{\downarrow}
\ema. \label{spinH}
\ee
The defining feature of altermagnetism is that these spin-up and spin-down sectors are interchanged under specific crystalline symmetries, such as $C_4,\ C_6$ rotations or mirrors. This spin exchange can be represented by two physically distinct operations: A. time-reversal, $T=\sigma_y\mathcal{K}$, where $\mathcal{K}$ is the complex conjugation, B. a $\pi$-rotation in spin space, $U_{\boldsymbol{n}_{\perp z} (\pi)}$, where the rotation axis is perpendicular to the $z$-direction. The general form of the latter is $U_{\boldsymbol{n}_{\perp z} (\pi)}=\cos \theta \sigma_x + \sin \theta \sigma_y$. For this spin-flipping operation, we fix the angle $\theta=0$ and define $U_\pi=\sigma_x$ for the remainder of this paper.
Although there are some exceptions of altermagnetism in non-collinear systems~\cite{Hayami2020,cheong_altermagnetism_2025,cheong_altermagnetism_2024,zhu_observation_2024} and the $p-$wave magnet violates Condition 5, we only focus on the conventional definition to identify altermagnetism. 

 We further note that, with SOC neglected (Condition 2), every crystalline operation acts trivially on spin, and the two spin sectors are exchanged only through the spin-flipping operations $T$ and $U_\pi$; the magnetic wallpaper and space groups identified as altermagnetic in this catalog are altermagnetic in this SOC-free (spin-group) sense. On the other hand, a magnetic group can also be treated with SOC, where every crystalline operation acts on spin through its spinful representation, and the same group can then describe a different magnetic order. For example, with SOC, the two-dimensional magnetic point group 2m$^{\prime}$m$^{\prime}$ describes a ferromagnet with the magnetic moment pointing out of the plane: the moment is preserved by the twofold rotation but reversed by the two spinful mirrors, so each mirror must be combined with time reversal, as the primes indicate.

\subsection{Spinless time reversal symmetry}

We assume these two operations ($T,\ U_\pi$) result in the same spin-flipping effect on the Hamiltonian,  although $T$ involves complex conjugation. Consequently, a global symmetry emerges:
\bee
SHS^{-1}=H, \label{spinless_symmetry}
\ee
where $S=U_\pi T$. Using the block-diagonal form of $H$, this implies $H_{\uparrow/\downarrow}^*=H_{\uparrow/\downarrow}$. In other words, the Hamiltonian block for each spin sector preserves its own time reversal symmetry, which we term \textit{spinless time-reversal symmetry}. In momentum space, this symmetry condition is expressed as:
\bee
H_\uparrow^*(\boldsymbol{k})=H_\uparrow(-\boldsymbol{k}),\ H_\downarrow^*(\boldsymbol{k})=H_\downarrow(-\boldsymbol{k}). \label{spinless_symmetry_k}
\ee
The symmetry physically means vanishing magnetic flux for each spin sector (zero phase in the Peierls substitution)\cite{Peierls1933} and each entry of the Hamiltonian in real space is real. This symmetry constraint ensures that the Hamiltonian satisfies Condition 5 and plays a pivotal role in realizing non-centrosymmetric altermagnets and the emergence of  nonsymmorphic altermagnetism, which will be discussed in later sections.

In the literature, the systematic classification of altermagnetism in point groups is primarily based on Laue groups, which inherently possess inversion symmetry \cite{Libor2022a,Libor2022b}. We note that the supplementary material of the same reference identifies the 3D magnetic point groups---both with and without inversion symmetry---that satisfy the condition criteria for conventional altermagnetism. This manuscript significantly expands upon that foundation by incorporating 2D magnetic point groups and extending the classification from point groups to wallpaper and space groups. Specifically, regardless of  inversion symmetry, we employ spinless time-reversal symmetry ( Eq.~\ref{spinless_symmetry}), which satisfies condition 5, to systematically identify a comprehensive set of altermagnetic candidates across both centrosymmetric and non-centrosymmetric systems.

Furthermore, nonsymmorphic symmetries, which involve translation operations, may enable novel spin configurations within the Brillouin zone. This specific class of nonsymmorphic altermagnetism is addressed immediately following the section on the 2D altermagnetism catalog.

\subsection{Collinear spin wallpaper/space group}

In the following sections, we catalog altermagnetism first within the 2D wallpaper groups and subsequently in the 3D space groups. To perform an exhaustive search, we begin with the framework of spin wallpaper groups (or spin space groups). Broadly, these spin groups are categorized into three types: collinear, coplanar, and non-coplanar \cite{LITVIN1974538,Litvin:a14103,Xiao2024,Jiang2024,Chen2024}. The conventional definition of altermagnetism falls strictly within the collinear category. In a collinear group, the essential symmetry element generating alternative spin configurations is an operation that acts independently on the spin and spatial degrees of freedom:
\begin{eqnarray}
[X\|R|\boldsymbol{v}] S_z(\boldsymbol{r})=\alpha S_z[\left(R|\boldsymbol{v}\right)^{-1} \boldsymbol{r}],
\end{eqnarray}
where the parameter $\alpha$ can be either 1 if $X=E$ or -1 if $X=U_\pi$, which indicates a composite spin-flipping operation. The first and second slots of $[\ldots\|\ldots]$ indicate spin operation and spatial operation, respectively, and the spatial operator $[R|\boldsymbol{v}]$ acts on atomic coordinates via rotation ($R$) and translation ($\boldsymbol{v}$). Given the collinear condition, the magnetic structure $S_z(\boldsymbol{r})$ is oriented solely along the $\pm z$-direction. A magnetic structure is considered invariant under a spin-space group symmetry if the operation satisfies $[X\|R|\boldsymbol{v}] S_z(\boldsymbol{r})=S_z(\boldsymbol{r})$. The set of all such crystalline symmetry elements with/without spin-flipping ($U_\pi$, $E$ ) forms a collinear spin wallpaper (or space) group.


\subsection{Type-III magnetic wallpaper/space group}

We now bridge collinear spin wallpaper/space groups to magnetic wallpaper/space groups. In magnetic groups, the time-reversal operation $T$ is the sole non-spatial operation. Using the spinless time-reversal symmetry defined in Eq.~(\ref{spinless_symmetry}), we invoke the equivalence between the time-reversal operation $T$ and the spin rotation $U_\pi$, allowing us to map magnetic wallpaper/space groups onto collinear spin wallpaper/space groups.
Since spin wallpaper/space groups describe magnetically ordered states locally, time-reversal symmetry is broken. Consequently, collinear spin groups correspond to Type-I, Type-III, and Type-IV magnetic groups (Type-II groups preserve time-reversal symmetry and are excluded).

Moreover, we can narrow our search for altermagnetism to a single type of magnetic group. First, Type-I groups describe ferromagnetism with spins aligned in a single direction and are therefore ruled out. Second, Type-IV groups must include a composite operation $T_a$ involving a spin-flip $U_\pi$ combined with a fractional translation $\boldsymbol{v}= 1/2$ (half the lattice constant of the magnetic unit cell). In momentum space, this operation takes the form:
\begin{equation}
    T_a=\bma
0 &  e^{-ik_\tau}\\
1 & 0
\ema \mathcal{K},
\end{equation}
where $e^{-ik_\tau}$ represents the phase factor associated with translation along the direction $\tau$ and $\mathcal{K}$ indicates the complex conjugation. We note that this composite operation doubles the size of the unit cell in the original wallpaper/space group. The symmetry $T_a^{-1}H(k)T_a=H(-k)$, together with Eq.~(\ref{spinless_symmetry_k}), imposes the constraint $H_\uparrow(\boldsymbol{k})=H_\downarrow(\boldsymbol{k})$ in violation of Condition 1. This spin degeneracy is characteristic of conventional antiferromagnetism rather than altermagnetism. Therefore, only Type-III groups can host altermagnetism. In the following, we focus on Type-III magnetic wallpaper/space groups to exhaustively identify altermagnets, employing the Opechowski-Guccione notation to label all possible candidates.

\subsection{Altermagnetic point group}

In Type-III magnetic groups, the spin-flipping operation is always coupled with crystalline operations, independent of spatial translations.
Consequently, translational symmetry does not fundamentally modify the symmetry criteria for altermagnetism.
We may therefore project the wallpaper or space groups onto their corresponding point groups, $\boldsymbol{R}$. After introducing magnetization, we
determine eligibility for altermagnetism in the set of the magnetic point groups that are reduced from type-III
magnetic wallpaper/space groups.
We define a magnetic point group that supports altermagnetism as an \textit{altermagnetic point group} (APG).
When translational symmetry is restored, a single APG may lift to single or multiple altermagnetic wallpaper groups (AWGs) or space groups (ASGs). The reason of multiple mappings is that distinct symmorphic and nonsymmorphic symmetries can be reduced to the same point-group symmetry in some cases.

We formally define the structure of an altermagnetic point group $\boldsymbol{M}$ as:
\begin{equation}
\boldsymbol{M} = [E \| \boldsymbol{H}] + \left[U_\pi \| \boldsymbol{R} \setminus \boldsymbol{H} \right],
\label{definition}
\end{equation}
where $\boldsymbol{H}$ is a subgroup of index $2$ in $\boldsymbol{R}$, and the complement  $\boldsymbol{R} \setminus \boldsymbol{H}$ represents the unique nontrivial coset of $\boldsymbol{H}$.
We refer to $\boldsymbol{H}$ as the \textit{halving subgroup} and $\boldsymbol{R} \setminus \boldsymbol{H}$ as the coset.
Physically, since $E$ is the identity operation, the spin degree of freedom remains invariant under operations in $\boldsymbol{H}$ and is flipped by operations in $\boldsymbol{R} \setminus \boldsymbol{H}$. This partition implies that the coset $\boldsymbol{R} \setminus \boldsymbol{H}$ possesses the same cardinality (size) as the subgroup $\boldsymbol{H}$. Consequently, exactly half of the operations in $\boldsymbol{R}$ preserve the spin (those in $\boldsymbol{H}$), while the other half (those in $\boldsymbol{R} \setminus \boldsymbol{H}$) flip it.
This halving structure guarantees zero net magnetization (Condition 4). Here is a simple example. The APG m$^{\prime}$ is denoted as $[E \| E] + \left[U_{\boldsymbol{n}_0}(\pi) \| m\right]$, where the prime symbol ($\prime$) indicates spin-reversal operation $U_{\boldsymbol{n}_0}(\pi)$.

In short, in the search for altermagnetism, in the following, we have to look for halving subgroups $\boldsymbol{H}$ satisfying Condition 4 and check whether $\boldsymbol{R} \setminus \boldsymbol{H}$ avoids spin degeneracy (Condition 1). The remaining conditions are automatically satisfied due to the spin block diagonalized Hamiltonian (Eq.~\ref{spinH}) and spinless time reversal symmetry (Eq.~\ref{spinless_symmetry_k}).

\subsection{Counting altermagnetism in the type-III magnetic space groups}\label{typeIII_count}

The type-III structure by itself already fixes most of the selection conditions, which permits a direct counting of the altermagnetic space groups before any group-by-group construction. In analogy with Eq.~(\ref{definition}), a type-III magnetic space group takes the form $[E \| \boldsymbol{H}] + [U_\pi \| \boldsymbol{G} \setminus \boldsymbol{H}]$, where $\boldsymbol{G}$ is the parent space group and the halving subgroup $\boldsymbol{H}$ carries index two. The spin-flip coset is thus guaranteed by the very definition of type III: the halving structure emerges naturally, the compensated magnetic order (Condition 4) is automatic. Only Condition 1 remains to be examined. It fails precisely when the coset contains an operation whose point-group part is the spatial inversion $P$: combining $[U_\pi \| P]$ with $S$ relates the two spin sectors at the same momentum and enforces $E_\uparrow(\boldsymbol{k})=E_\downarrow(\boldsymbol{k})$ throughout the BZ---the momentum-space fingerprint of the combined time-reversal--inversion symmetry. A type-III magnetic space group therefore describes a conventional antiferromagnet when the inversion resides in the spin-flip coset $\boldsymbol{G} \setminus \boldsymbol{H}$, and an altermagnet otherwise, i.e., when the inversion belongs to the halving subgroup $\boldsymbol{H}$ or is absent from $\boldsymbol{G}$ altogether. Among the 674 type-III magnetic space groups~\cite{LitvinTables2013}, we count 252 whose coset contains the inversion; these are conventional antiferromagnets with spin-degenerate bands at every momentum and they constitute the prototype SST-1 of Ref.~\cite{Yuan2021PRM}. The remaining $674-252=422$ type-III magnetic space groups are altermagnetic.

Although the type-III route requires only this single condition, checking 674 magnetic space groups one by one is impractical, and the count alone reveals nothing about the spin distribution in the BZ. In the following sections we therefore proceed constructively from the point groups: we identify the altermagnetic point groups through their halving subgroups and then restore the translational symmetries to generate the altermagnetic wallpaper and space groups, which reproduces the number 422 and, moreover, yields the momentum-space spin texture of each group.

\begin{table*}
\renewcommand\arraystretch{2.5}
\setlength{\tabcolsep}{1.8mm}{
    \begin{tabular}{|*{11}{c|}}
        \hline
        Crystal systems & \multicolumn{2}{c|}{Monoclinic} &  \multicolumn{2}{c|}{Orthorhombic} & \multicolumn{2}{c|}{Square} &  \multicolumn{4}{c|}{Hexagonal} \\  \cline{1-11}
         Point group $\boldsymbol{R}$ &  $\boldsymbol{C}{_{1}}$ &  $\boldsymbol{C}{_{2}}$ &  {\clr $\boldsymbol{C}{_{1h}}$} &  $\boldsymbol{C}{_{2v}}$ & $\boldsymbol{C}{_{4}}$ &  $\boldsymbol{C}{_{4v}}$ &  $\boldsymbol{C}{_{3}}$ &  {\clr $\boldsymbol{C}{_{3v}}$}  &  $\boldsymbol{C}{_{6}}$ &  $\boldsymbol{C}{_{6v}}$  \\ \hline
          Wallpaper group  &  $\#1$ &  $\#2$ &  {\clr $\#3-5$ } &  $\#6-9$ & $\#10$ &  $\#11-12$ &  $\#13$ &  {\clr  $\#14-15$} &  $\#16$ &  $\#17$ \\ \hline

        Halving Subgroup $\boldsymbol{H}$ & $\times$ & $\times$ & {\clr $\boldsymbol{C}{_{1}}$} & $\boldsymbol{C}{_{2}}$ & $\boldsymbol{C}{_{2}}$ & $\boldsymbol{C}{_{4}}$/$\boldsymbol{C}{_{2v}}$ & $\times$ & {\clr $\boldsymbol{C}{_{3}}$} & $\times$ & $\boldsymbol{C}{_{6}}$ \\ \hline

         APG & $\times$ & $\times$ & {\clr m$^{\prime}$} & 2m$^{\prime}$m$^{\prime}$ & 4$^{\prime}$ & 4m$^{\prime}$m$^{\prime}$, 4$^{\prime}$m$^{\prime}$m & $\times$ & {\clr 3m$^{\prime}$} & $\times$ & 6m$^{\prime}$m$^{\prime}$ \\ \hline

          AWG & $\times$ & $\times$ & {\clr \makecell[c]{ P1m$^{\prime}$1 \\ P1g$^{\prime}$1 \\ C1m$^{\prime}$1 }} & \makecell[c]{ P2m$^{\prime}$m$^{\prime}$,~P2m$^{\prime}$g$^{\prime}$ \\ P2g$^{\prime}$g$^{\prime}$,~C2m$^{\prime}$m$^{\prime}$ } & P4$^{\prime}$ & \makecell[c]{ P4m$^{\prime}$m$^{\prime}$,~P4$^{\prime}$mm$^{\prime}$ \\ P4$^{\prime}$m$^{\prime}$m,~P4g$^{\prime}$m$^{\prime}$ \\ P4$^{\prime}$gm$^{\prime}$,~P4$^{\prime}$g$^{\prime}$m}  & $\times$ & {\clr \makecell[c]{ P3m$^{\prime}$1 \\ P31m$^{\prime}$ }}  & $\times$ & P6m$^{\prime}$m$^{\prime}$ \\ \hline

    \end{tabular}}
    \caption{Classification of 2D point groups and their associated altermagnetic groups. The table lists the crystal system, point group, halving subgroup ($\boldsymbol{H}$), Altermagnetic Point Group (APG), and the resulting Altermagnetic Wallpaper Groups (AWG). Non-centrosymmetric groups are highlighted in red, while point groups incompatible with altermagnetism are marked with `X'. For clarity, APGs and AWGs are labeled using Opechowski-Guccione notation.}
    \label{2D Point group}
\end{table*}

\section{Catalog of 2D altermagnetic wallpaper groups}

We systematically analyze the magnetic extension of the ten 2D point groups to identify APGs that can support altermagnetic order, and we construct the associated AWGs. Our analysis identifies 17 distinct AWGs, comprising 12 centrosymmetric and 5 non-centrosymmetric groups. We further discuss their characteristic spin state distributions and symmetry features within the BZ.

We begin by briefly reviewing 2D lattice systems. Two-dimensional crystal symmetries are classified into ten distinct point groups, denoted as $\boldsymbol{C}_{n}$ and $\boldsymbol{C}_{nv}$ (where $n = 1, 2, 3, 4, 6$). Each point group is generated by rotation symmetry $C_n$ and/or reflection symmetry $M$. Specifically, $\langle C_n \rangle$ generates the cyclic groups $\boldsymbol{C}_{n}$, while $\langle M,C_n \rangle$ generates the dihedral groups $\boldsymbol{C}_{nv}$. Furthermore, incorporating collinear magnetic order into the 2D lattice gives rise to 11 type-III magnetic point groups and 26 type-III magnetic wallpaper groups. However, since not all magnetic groups can accommodate altermagnetic order, a detailed symmetry-based analysis is required.

Building on the framework established in the previous section, it suffices to examine two conditions to determine altermagnetism: compensated collinear magnetic order (Condition 4) and momentum-dependent spin splitting (Condition 1). While conventional antiferromagnetism satisfies the first condition, it is distinguished by the presence of spin degeneracy. Our approach to screening APGs based on these two criteria is detailed as follows.

\begin{itemize}
\item{\textbf{Compensated collinear magnetic order}: Point groups of altermagnetic systems must include a halving subgroup $\boldsymbol{H}$ that connects the same-spin sublattice. The coset $\boldsymbol{R} \setminus \boldsymbol{H}$ associated with this halving subgroup must link sublattices with opposite spins, ensuring zero net magnetization. Consequently, point groups lacking such halving subgroups are incompatible with altermagnetism. For instance, a halving group does not exist in the group $\boldsymbol{C}_{3}$, which connects sublattices with the same spin only due to inherent geometric frustration in the spin arrangement.}

\item{\textbf{Breaking $C_{2z}T$ symmetry}: The combined symmetry $C_{2z}T$, involving inversion and time reversal, enforces Kramers degeneracy, resulting in spin-degenerate electronic states at every momentum. To preclude the degeneracy characteristic of antiferromagnetism, opposite-spin sublattices must not be connected by the $C_{2z}$ rotation. In other words, if the point group includes inversion symmetry, the inversion operation must reside within the halving subgroup ($C_{2z} \in \boldsymbol{H}$) rather than in the coset ($C_{2z} \in \boldsymbol{R} \setminus \boldsymbol{H}$).}
\end{itemize}

\begin{figure}
\centerline{\includegraphics[width=0.65\textwidth]{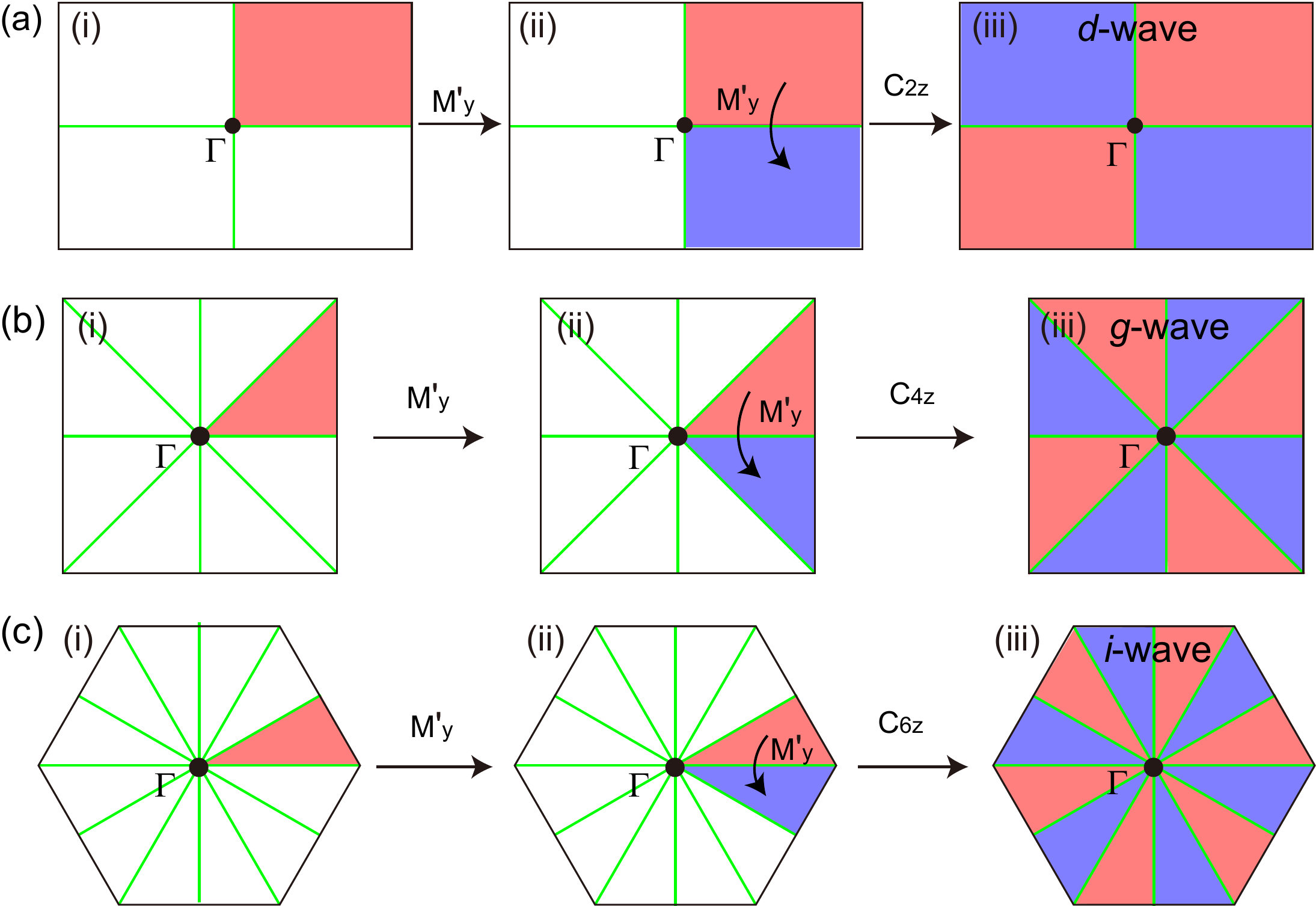}}
\caption{ (color online) Stepwise construction of spin-momentum textures in the Brillouin zone for representative AWGs: (a) $P2m^{\prime}m^{\prime}$, (b) $P4m^{\prime}m^{\prime}$, and (c) $P6m^{\prime}m^{\prime}$. Red and blue regions denote spin-up and spin-down polarization, respectively. The panels illustrate the generation process: (i) The initial spin-polarized irreducible BZ (red). (ii) The texture generated by applying the spin-flipping mirror generator $M_y^{\prime}$ (mapping red to blue). (iii) The complete spin-splitting pattern obtained by applying the rotation generator $C_{nz}$. Green lines indicate spin-degenerate nodal lines protected by mirror symmetries.}
\label{type}
\end{figure}

By evaluating the five generator types ($M, C_{2z}, C_{3z}, C_{4z}, C_{6z}$), we find that only mirror symmetry $M$ and $C_{4z}$ rotational symmetry are permitted in the coset $\boldsymbol{R} \setminus \boldsymbol{H}$ to connect opposite-spin sublattices. Hence, when $C_{nz}$ rotational symmetries ($n = 2, 3, 6$) are preserved, they must belong to the halving subgroup $\boldsymbol{H}$. Applying these requirements alongside the APG structure defined in Eq.~\ref{definition}, we systematically examine all ten 2D point groups to determine their eligibility to host altermagnetic order. The point groups $\boldsymbol{C}_{1}$ and $\boldsymbol{C}_{3}$ lack a halving subgroup and therefore cannot generate APGs. Furthermore, although $\boldsymbol{C}_{2}$ and $\boldsymbol{C}_{6}$ possess halving subgroups, these magnetic point groups include $C_{2z}T$ symmetry, rendering them compatible only with antiferromagnetism. In contrast, the remaining six point groups ($\boldsymbol{C}_{1h}$, $\boldsymbol{C}_{2v}$, $\boldsymbol{C}_{3v}$, $\boldsymbol{C}_{4}$, $\boldsymbol{C}_{4v}$, $\boldsymbol{C}_{6v}$) contain halving subgroups that satisfy the symmetry criteria for altermagnetic order. By applying Eq.~\ref{definition} to these six point groups, we identify seven APGs. Notably, the point group $\boldsymbol{C}_{4v}$ contains two distinct halving subgroups, $\boldsymbol{C}_{4}$ and $\boldsymbol{C}_{2v}$, both of which satisfy the necessary criteria. Selecting $\boldsymbol{C}_{4}$ as the halving subgroup yields the APG 4m$^{\prime}$m$^{\prime}$, while selecting $\boldsymbol{C}_{2v}$ leads to 4$^{\prime}$m$^{\prime}$m. Each of the remaining five point groups generates a single corresponding APG. By incorporating translational symmetries and composite operations involving translations (such as glide reflections), these APGs give rise to 17 distinct AWGs (a number coincidentally identical to the total count of non-magnetic wallpaper groups), comprising 12 centrosymmetric and 5 non-centrosymmetric cases, as shown in Table~\ref{2D Point group}.

Non-centrosymmetric lattices can host altermagnetic order arising from spinless time-reversal symmetry (Eq.~\ref{spinless_symmetry}) with symmetry operator $S$. This symmetry functions analogously to inversion symmetry in momentum space by enforcing the equivalence of spin states at opposite momenta (Condition 5). Consequently, the non-centrosymmetric altermagnetic point groups (APGs) $m^{\prime}$ and $3m^{\prime}$ become symmetry-equivalent to their centrosymmetric counterparts. This equivalence in momentum space is expressed as:
\begin{equation}
    m^{\prime}\times \widetilde{2} \cong 2m^{\prime}m^{\prime}, \quad 3m^{\prime}\times \widetilde{2} \cong 6m^{\prime}m^{\prime},
    \label{equal}
\end{equation}
where $\widetilde{2}=\{ E, S \}$ denotes the group generated by the effective inversion $S$, distinct from the standard twofold rotation $2$ in the International notation. For instance, in the group $m^{\prime}$ (containing $M^{\prime}_{x}$), the combination with $S$ generates a new effective mirror symmetry $\bar{M}^{\prime}_{y} = S M^{\prime}_{x}$. The resulting extended symmetry set $\{ E, M^{\prime}_{x}, S, \tilde{M}^{\prime}_{y}\}$ is isomorphic to the magnetic point group $2m^{\prime}m^{\prime}$. Similarly, $3m^{\prime}$ is effectively promoted to $6m^{\prime}m^{\prime}$.

We note that for APGs already possessing $C_{2z}$ rotational symmetry, the inclusion of $S$ imposes no additional constraints, as $S$ acts equivalently to $C_{2z}$ in mapping the Brillouin zone ($\boldsymbol{k} \rightarrow -\boldsymbol{k}$). Moreover, we emphasize that while $C_{2z}$ and $S$ share this momentum-space equivalence, they remain fundamentally distinct symmetries in real space.

Once an AWG is identified, we map its spin configuration in the BZ as a spin-momentum-locked band structure. Regardless of the specific BZ geometry, the existing literature categorizes these configurations into three primary types: $d$-wave, $g$-wave, and $i$-wave~\cite{Libor2022a,Libor2022b}. These classifications can be understood via an expansion in real spherical harmonics $Y_{\ell}^m(\theta,\phi)$. For a two-dimensional system ($\theta=\pi/2$, setting $m=\ell$), the spin-splitting energy of the low-energy Hamiltonian near the $\Gamma$ point is expressed as:
\begin{align}
    \Delta E_{d\text{-wave}} \propto& \left[ k^2Y_2^2(\pi/2,\phi) + \mathcal{O}\left(k^4Y_4^4(\pi/2,\phi)\right) \right] \sigma_z,\label{d-wave} \\
    \Delta E_{g\text{-wave}} \propto& \left[ k^4Y_4^4(\pi/2,\phi) + \mathcal{O}\left(k^8 Y_8^8(\pi/2,\phi)\right) \right] \sigma_z, \label{g-wave} \\
    \Delta E_{i\text{-wave}} \propto& \left[ k^6Y_6^6(\pi/2,\phi) + \mathcal{O}\left(k^{12}Y_{12}^{12}(\pi/2,\phi)\right) \right] \sigma_z, \label{i-wave}
\end{align}
where $\sigma_z$ denotes the spin energy splitting and $\phi$ is the azimuthal angle relative to the $\Gamma$ point. Specifically in the 2D limit, the parameter $2\ell$ corresponds directly to the number of alternating spin-polarized sectors, which equals the number of spin-degenerate nodal lines. Due to rotation, mirror, and spinless time-reversal symmetries, these spin sectors are distributed evenly with an angular periodicity of $\pi/\ell$. (We reserve the discussion of the 3D correspondence between $Y_{\ell}^m(\theta,\phi)$ and 3D spin regions for later.)

To systematically classify how altermagnetic orders manifest the harmonic degree $\ell$ within the BZ, we establish the following three-step analysis, using the centrosymmetric AWGs P2m$^\prime$m$^\prime$, P4m$^\prime$m$^\prime$, and P6m$^\prime$m$^\prime$ as illustrative examples:

\begin{enumerate}
    \item[(a)] \textbf{Identify Generators:} Determine the symmetry generators of the AWG required to construct the spin texture. For instance, for the groups P2m$^\prime$m$^\prime$, P4m$^\prime$m$^\prime$, and P6m$^\prime$m$^\prime$, the generators $M^{\prime}_{y}$ (mirror) and $C_{nz}$ (rotation) are given.

    \item[(b)] \textbf{Construct Spin Texture:} We start with spin up in the initial irreducible wedge of the BZ. The full texture is then generated by applying the generators (see Fig.~\ref{type}). First, as shown in Fig.~\ref{type}(a-i), (b-i), and (c-i), we define the initial spin-up distribution (red). The generator $M^{\prime}_{y}$ maps this region to an adjacent sector with spin down (blue). Finally, applying the rotation generator $C_{nz}$ ($n=2,4,6$) extends these sectors across the full BZ, yielding the complete spin-splitting configuration shown in Fig.~\ref{type}.

    \item[(c)] \textbf{Classify Harmonic Degree $\ell$:} We analyze the resulting spin-splitting configuration to determine the altermagnetic wave type. A fourfold pattern ($\ell=2$) corresponds to $d$-wave, eightfold ($\ell=4$) to $g$-wave, and twelvefold ($\ell=6$) to $i$-wave. It is crucial to note that the $n$-fold symmetry of the spin distribution is not necessarily equivalent to the crystallographic $C_n$ rotation symmetry. For example, in Fig.~\ref{type}, P2m$^\prime$m$^\prime$ displays four distinct sectors near $\Gamma$ ($d$-wave). Similarly, P4m$^\prime$m$^\prime$ and P6m$^\prime$m$^\prime$ exhibit eight and twelve separate spin regions, respectively, categorizing them as $g$-wave and $i$-wave altermagnets.
\end{enumerate}

\begin{figure}
\centerline{\includegraphics[width=0.8\textwidth]{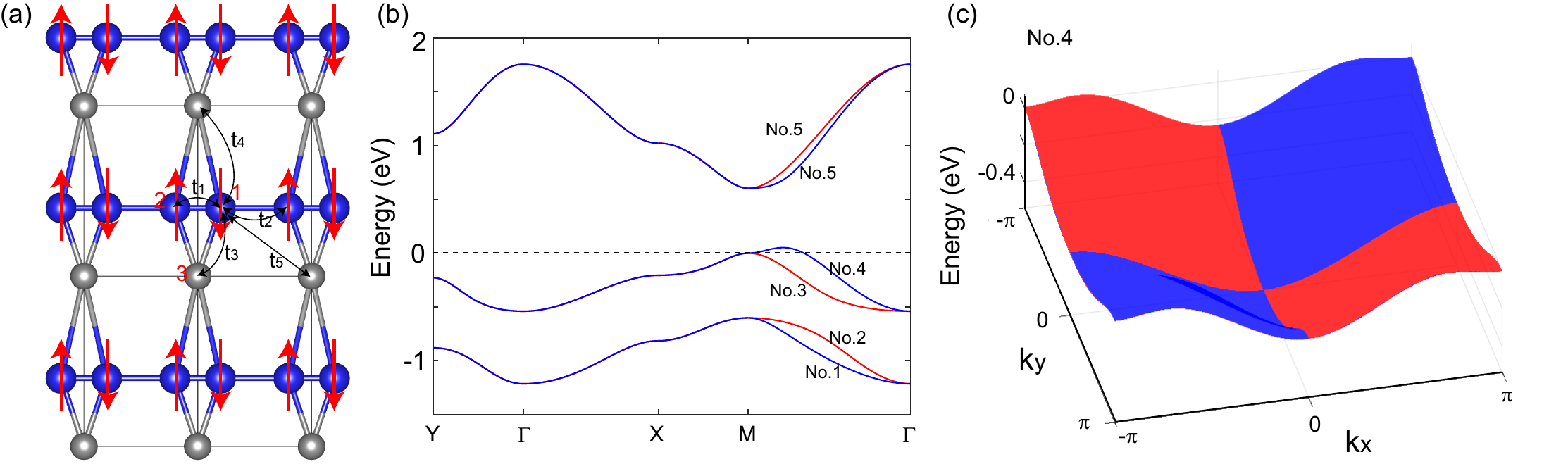}}
\caption{(color online) Crystal lattice and band structure for AWG P1m$^{\prime}$1.  (a) Two-dimensional altermagnetic crystal lattice, where No.1 and 2 carry spin-down and spin-up moments, and No.3 is nonmagnetic atom. (b) Band structure along high symmetry lines. Red and blue lines are spin up and down bands. (c) Three-dimensional energy dispersion of the 4th band over the Brillouin zone.
\label{WG3} }
\end{figure}

By using spinless time-reversal symmetry, the non-centrosymmetric AWGs P1m$^{\prime}$1, P3m$^{\prime}$1, and P31m$^{\prime}$ exhibit $d$-wave, $i$-wave, and $i$-wave altermagnetic orders, respectively.

We use a simple reflection symmetry to show that a $d$-wave altermagnetism can emerge in AWG P1m$^{\prime}$1. Fig.~\ref{WG3}(a) displays the corresponding two-dimensional crystal lattice, containing both magnetic and nonmagnetic atoms. The construction of the tight-binding model is based on this structure  (details in Supplemental Material (SM)).   The spin-flipping mirror $M_{x}^{\prime}$ links the  opposite-spin sites. Combined with spinless time-reversal symmetry $S$, this yields an effective spin-flipping mirror $\bar{M}^{\prime}_{y}=SM_{x}^{\prime}$. These two mirror symmetries lead to the spin up and down degeneracy along the x- and y-axes, as evident in the band structure shown in  Fig.~\ref{WG3}(b). Moreover, the three-dimensional energy dispersion of the fourth band over the full Brillouin zone [Fig.~\ref{WG3}(c)] clearly displays a fourfold pattern, providing direct evidence that P1m$^{\prime}$1 realizes a $d$-wave altermagnet.

\begin{table}
\centerline{\includegraphics[width=0.8\textwidth]{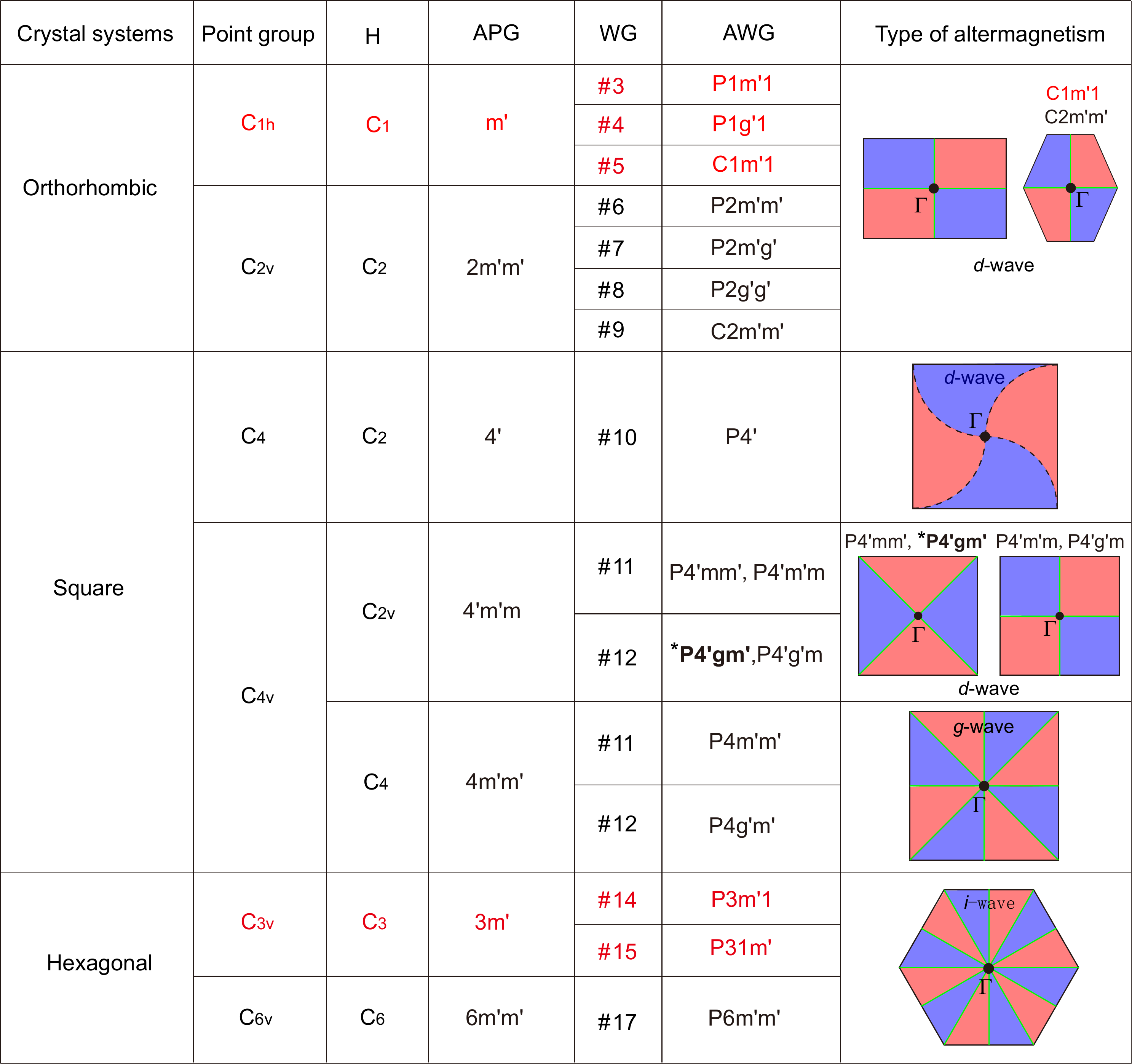}}
\caption{(color online) Catalog of altermagnetic orders for all Altermagnetic Wallpaper Groups (AWGs) and their corresponding spin-splitting distributions in the primitive Brillouin zone. The table categorizes groups by crystal system, point group, and altermagnetic wave type ($d$-wave, $g$-wave, $i$-wave). Non-centrosymmetric groups are highlighted in red text. In the BZ diagrams, red and blue sectors represent spin-up and spin-down polarizations, respectively. Green solid lines denote symmetry-enforced nodal lines that are fixed in momentum space, whereas black dashed lines represent accidental nodal lines not protected by lattice symmetries, which may shift. Label $^*$ on P4$^\prime$gm$^\prime$ to indicate the special nonsymmorphic BZ.
\label{2D} }
\end{table}

Incorporating translational symmetry reveals further distinctions in the primitive-cell BZ spin distributions. Specifically, two distinct AWGs belonging to the same APG can exhibit different spin-splitting configurations. The sole example in 2D involves the pairs P4$^{\prime}$m$^{\prime}$m / P4$^{\prime}$g$^{\prime}$m and P4$^{\prime}$mm$^{\prime}$ / P4$^{\prime}$gm$^{\prime}$. While all exhibit $d$-wave character, the symmetries $m^{\prime}$ (combining mirror and spin-flip) enforce different spin-splitting placements. For P4$^{\prime}$m$^{\prime}$m (and P4$^{\prime}$g$^{\prime}$m), the spin-degenerate nodal lines align with the $k_x/k_y$ axes and zone boundaries. In contrast, for P4$^{\prime}$mm$^{\prime}$ (and P4$^{\prime}$gm$^{\prime}$), these nodes align along the BZ diagonals (Table~\ref{2D}). Notably, as detailed in the next section, P4$^{\prime}$gm$^{\prime}$ represents the only 2D case where a nonsymmorphic symmetry drives a new altermagnetic spin-splitting distribution by eliminating zone-boundary nodal lines.

Table~\ref{2D} summarizes the catalog of altermagnetic orders for all AWGs and their primitive-cell BZ spin distributions (non-centrosymmetric groups highlighted in red). Red and blue sectors denote spin-up and spin-down, separated by spin-degenerate nodal lines. Green solid lines indicate nodes locked by mirror symmetries (fixed in momentum space), while black dashed lines represent nodes unlocked by lattice symmetries, which may shift because the spin-flipping rotation $C_{4z}U_\pi$ cannot fix the nodes. In other words, except for P4$^\prime$, the remaining AWGs exhibit spin-locking feature. Only the spin distribution in P4$^\prime$ can vary by symmetry-preserving perturbations. Table~\ref{2D} depicts the minimal spin textures guaranteed by symmetry. In realistic materials, accidental crossings between multiple bands with opposite spins may introduce additional spin alternations, complicating the distribution. In principle, however, such complex band structures can be decomposed into superpositions of the elementary two-band models described in Eqs.~\ref{d-wave}--\ref{i-wave} within relevant energy windows (see Eq.~\ref{nonsHM} for an example).


We address the orthorhombic system, which allows for rectangular or centered-rectangular Bravais lattices. While both yield $d$-wave altermagnetism, their BZ geometries differ: rectangular lattices utilize a standard rectangular BZ, whereas centered rectangular lattices (e.g., C1m$^{\prime}$1 and C2m$^{\prime}$m$^{\prime}$) possess a primitive BZ that is reminiscent of the hexagonal cell.

The above construction provides a complete 2D catalog of altermagnetic orders in all AWGs and their characteristic spin-momentum textures in the primitive-cell BZs. The catalog sets the stage for the nonsymmorphic altermagnetism and 3D generalizations discussed in the following sections.

 \section{Nonsymmorphic Altermagnetism}

 The altermagnets cataloged in the preceding sections all fit within the conventional classification introduced earlier. This section turns to a richer story: certain crystal symmetries can give rise to a class of altermagnetism that lies outside that conventional picture. We first walk through the only two-dimensional example in which this new class is realized (Sec.~\ref{2D_nonsym}), and then illustrate how the same phenomenon extends to three dimensions through two representative cases (Sec.~\ref{3D_nonsym}).

 \subsection{The only magnetic wallpaper group with nonsymmorphic altermagnetism}\label{2D_nonsym}

 We examine the nonsymmorphic characteristics of the altermagnetic wallpaper group $P4'gm'$.
Except for translations, the altermagnetic group is generated by three symmetry operations: $C_{4z}T$, $M_{x=y}T$, and the glide mirror $M_x$.
The first two involve time-reversal $T$, which is equivalent to flipping the spin.
The defining nonsymmorphic feature is the spin-conserving glide mirror $M_x$, which involves a reflection $x \to -x$ followed by a half lattice constant translation in the $y$ direction.
In momentum space (assuming lattice constant $a_y=1$), this operator acts as:
\begin{equation}
M_x: (k_x,k_y) \rightarrow (-k_x,k_y)e^{ik_y/2}.
\end{equation}
This phase factor reflects the system's invariance under a mirror operation combined with a vertical shift of half a lattice constant. Recall the definition of the spinless time-reversal symmetry, represented by the antiunitary operator $S=TU_\pi$.
Under $S$, the spin sectors are decoupled (not exchanged), and the momentum transforms as:
\begin{equation}
S:(k_x,k_y) \rightarrow -(k_x,k_y)\mathcal{K}.
\end{equation}
We now construct the composite antiunitary operator $M_xS$.
Since $S$ acts locally in real space, the fractional translation inherent to the glide persists in the combined operator:
\begin{equation}
M_xS : (k_x,k_y) \rightarrow (k_x,-k_y)e^{ik_y/2}\mathcal{K}.
\end{equation}
Squaring this combined operator yields a full lattice translation along the $y$-direction: $(M_xS)^2=e^{ik_y}$.
Crucially, along the BZ boundary at $k_y=\pi$, we find $(M_xS)^2 = -1$.
Since $M_xS$ is antiunitary, this relation enforces a Kramers-like twofold degeneracy within each spin sector along the $k_y=\pi$ line.

\begin{figure}
\centerline{\includegraphics[width=0.8\textwidth]{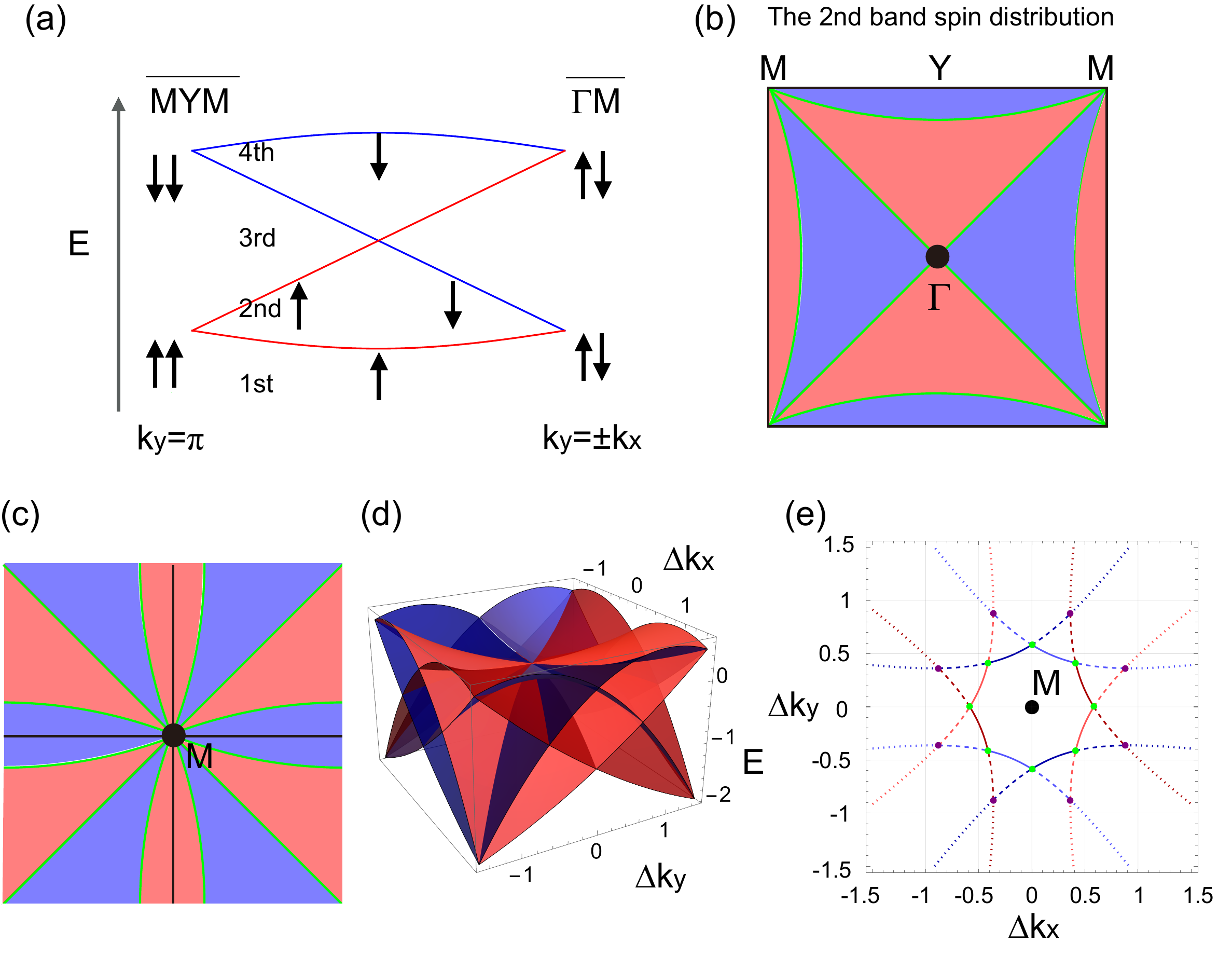}}
\caption{(color online)
\textbf{Nonsymmorphic altermagnetism in the magnetic wallpaper group $P4'gm'$.}
(a) Schematic of the hourglass dispersion along a path connecting the BZ boundary ($k_y=\pi$) to the diagonal symmetry lines ($k_x = \pm k_y$). A symmetry-protected crossing (nodal line) occurs between the 2nd and 3rd bands with opposite-spin polarizations.
(b) Spin distribution of the 2nd occupied band over the BZ, while the 3rd band exhibits the opposite-spin distribution (not shown).
(c) Detailed spin texture of the 2nd band near the $M$ point, revealing 12 nodal lines (green) where the spin character inverts as the path encircles $M$.
(d) Three-dimensional energy dispersion of the effective four-band Hamiltonian near the $M$ point  ($h(k)\equiv-k^2/6$, $A\equiv1$), showing the four intertwined spin-split bands.
(e) Fermi surface topology at $E=-0.15$ derived from the $M$-point effective model in the axes of $\Delta k_x=k_x - \pi $ and $\Delta k_y=k_y - \pi $. The green dots indicate crossing points between the 1st and 2nd bands (inner ring), while the purple dots mark crossings between the 2nd and 3rd bands (outer ring). While the solid curves and the dotted curves represent the 1st and 3rd lowest bands, respectively, the dashed curves (2nd band) highlight the emergent, albeit extrinsic, $i$-wave-like feature: the spin polarization along the dashed Fermi surface changes sign 12 times, matching the counting of an $i$-wave altermagnet.
}
\label{nonsymmorphic}
\end{figure}

Generally, the spin-up and spin-down doublets are split in energy along $k_y=\pi$, except at the high-symmetry $M$ point where a fourfold degeneracy is protected.
Conversely, along the diagonal lines ($k_x=\pm k_y$), the system exhibits a different twofold degeneracy of spin-up and spin-down states.
To satisfy these distinct degeneracy constraints, the spectrum along a path connecting $k_y=\pi$ to the diagonal lines must exhibit an hourglass dispersion\cite{wang_hourglass_2016,Sato2016}.
As illustrated in Fig.~\ref{nonsymmorphic}(a), along a trajectory from the $k_y=\pi$ line to the diagonal axes ($\overline{\Gamma M}$), the second and third bands possessing opposite-spin characters are forced to cross. The crossing of the second and third bands forms a nodal line intersecting the $M$ point, located between the zone boundary and the diagonal axes.

By applying $C_{4z}T$ symmetry, this local spectral feature extends to the entire BZ.
While the first and fourth occupied bands maintain the spin-splitting configuration depicted in Table~\ref{2D} for P4'gm', the intermediate bands (2nd and 3rd) display a complex spin texture as shown in Fig.~\ref{nonsymmorphic}(b).

We first analyze the low-momentum physics near the $\Gamma$ point.
For a four-band model, the effective Hamiltonian with momentum slightly displaced from $\Gamma$ can be approximated as two decoupled blocks:
\begin{equation}
H_\Gamma(\boldsymbol{k}) \approx \begin{pmatrix}
A Y_2(k,\phi) \sigma_z + f(k)\sigma_0 & 0 \\
0 & B Y_2(k,\phi) \sigma_z + g(k)\sigma_0
\end{pmatrix},
\end{equation}
defined in the ordered basis of $(\uparrow, \downarrow, \uparrow, \downarrow)$.
Here, $A$ and $B$ are material-dependent constants that determine the strength of the $d$-wave splitting, $(k,\phi)$ denotes polar coordinates in momentum space, and the $d$-wave harmonic is defined as $Y_2(k,\phi) \equiv k^2 Y^2_2(\pi/2,\phi)$.
We assume $f(k) \ll g(k)$, ensuring that the two sets of the spin-split bands remain energetically separated without crossing at low momentum.
This limit essentially recovers the conventional $d$-wave altermagnetism described by the two-band model in the previous section (Eq.~\ref{d-wave}).

Much richer physics emerges at the $M$ point.
As a path encircles $M$, the two intermediate (2nd and 3rd) bands exhibit a spin-character inversion 12 times, corresponding to 12 spin-degenerate nodal lines [see Fig.~\ref{nonsymmorphic}(c)].
These nodal lines arise from crossings between different pairs of bands.
For instance, considering the second lowest band: the nodal lines along the diagonal directions ($\overline{\Gamma M}$) correspond to spin-flipping crossings with the first (lowest) band.
Conversely, the remaining eight spin-flipping lines stem from crossings between the second and third bands.
Since the third and fourth bands also exchange spin character at the crossings along $k_x=\pm k_y$, and $M$ is a fourfold degenerate point, a minimum four-band model is required to capture the low-momentum expansion.
Given the $C_{4z}T$ symmetry, the harmonic functions with $l=2, m=2$ continue to describe the lowest order of the spin splitting.
The effective four-band Hamiltonian near $M$ takes the form:
\begin{equation}
H_M(k,\phi) = h(k)\mathbf{I}_{4\times 4} + A \cdot \text{diag}\left( Y_2(k,\phi+\theta_0), -Y_2(k,\phi+\theta_0), Y_2(k,\phi-\theta_0), -Y_2(k,\phi-\theta_0) \right), \label{nonsHM}
\end{equation}
in the same basis as $H_\Gamma$.
Here, $h(k)$ is the global dispersion function and $\theta_0 \in (0, \pi/4)$ is a phase offset.
The effective Hamiltonian preserves the three generators of the altermagnetic wallpaper group:
\begin{align}
C_{4z}T H_M (k,\phi+\pi/2) (C_{4z}T)^{-1} &= H_M (k,\phi), \\
M_{x=y}T H_M (k,-\phi+\pi/2) (M_{x=y}T)^{-1} &= H_M (k,\phi), \\
M_x H_M (k,-\phi+\pi) M_x^{-1} &= H_M (k,\phi),
\end{align}
where the operators are represented as $C_{4z}T= \tau_0\sigma_x$, $M_{x=y}T= \tau_x\sigma_x$, and $M_x= \tau_x\sigma_0e^{ik_y/2}$.
This minimal model reveals that the two sets of two-band spin-splittings (represented by the diagonal blocks) are locked at the same energy level but rotated by $2\theta_0$.
The $M_x$ symmetry specifically forbids the energetic separation of these blocks.
Consequently, the four spin-split bands (red $\uparrow$, blue $\downarrow$) are forced to intertwine, as shown in the energy dispersion in Fig.~\ref{nonsymmorphic}(d).

To confirm that these effective Hamiltonians ($H_\Gamma$, $H_M$) faithfully capture the physics of the nonsymmorphic group $P4'gm'$, we construct an eight-band tight-binding model (details in SM).
The tight-binding results (Fig.~S1 in the SM) are clearly consistent with the two separated $d$-wave splittings near $\Gamma$ and the four intertwined bands near $M$ from the effective models derived above.

Finally, we analyze the Fermi surface topology derived from the $M$-point effective Hamiltonian.
At a fixed energy, the Fermi surface exhibits 12 spin-crossing points formed by the intersection of curves with opposite-spin polarization (Fig.~\ref{nonsymmorphic}(e)).
These crossings manifest the 12 distinct spin regions of the second band discussed earlier.
While the 12 sign changes of the spin polarization around this Fermi surface superficially resemble $i$-wave altermagnetism (dashed curve in panel e), the underlying origin is fundamentally different.
This $i$-wave-like character is an emergent property of two sets of spin-split bands ($Y_2(k,\phi+\theta_0)\sigma_z,\ Y_2(k,\phi-\theta_0)\sigma_z$) coexisting at the same energy level, rather than a single pair of bands with high-order harmonics.
Specifically, the crossing points on the inner ring (green dots) arise from the intersection of the first and second bands, while those on the outer ring (purple dots) stem from the second and third bands.
Thus, the full Fermi surface structure reflects the complex interference of two iso-energetic $d$-wave splittings, distinct from the simple single pair of splitting bands of conventional altermagnetism.

Returning to the classification of 2D altermagnetism in Table~\ref{2D}, we identify $ P4'gm'$ as the only 2D group exhibiting symmetry-enforced band crossings accompanied by spin exchange as part of the hourglass dispersion.
This feature results in a novel spin texture within the BZ, which we designate with a star ($*$) in Table~\ref{2D}. 
With and without nonsymmorphic symmetries, near the $\Gamma$ point, the low-momentum physics remains well-described by a conventional two-band model governed by standard spherical harmonics ($d$-wave [Eq.~\ref{d-wave}]).
However, $P4'gm'$ is distinct in that its physics near the $M$ point requires a description involving two identical $d$-wave two-band models, rotationally offset from one another. While these nonsymmorphic symmetries introduce additional spin exchanges, they do not fundamentally alter the symmetry of the lowest-order harmonics.


\begin{table}
\centering
\caption{List of space groups compatible with altermagnetic order that exhibit screw axis $\tilde{C}_{2}$ and glide plane $\tilde{M}$. The combination of these symmetries with the $S$ symmetry can enforce same-spin degenerate lines or planes. Notably, only space groups Nos.~76, 78, 85, and 86 host unlocked opposite-spin degenerate planes.}
\renewcommand{\arraystretch}{2.0}
\setlength{\tabcolsep}{7mm}
\begin{tabular}{lcc}
\toprule
\text{Crystal System} & \text{Screw Axis $\tilde{C}_{2}$} & \text{Glide Plane $\tilde{M}$} \\
\midrule
Orthorhombic & \begin{tabular}[c]{@{}l@{}}17--20, 26, 29, 31, 33, 36, \\ 51--64, 67, 68\end{tabular} &  \\
\cmidrule{2-3}
Tetragonal & \begin{tabular}[c]{@{}l@{}}76, 78, 90--92, 94--96, \\ 112--114, 127--130, 135--138\end{tabular} & \begin{tabular}[c]{@{}l@{}}85, 86, 100--106, 108--110, 112--118, \\ 122, 124--138, 141, 142\end{tabular} \\
\cmidrule{2-3}
Hexagonal \& Cubic & \begin{tabular}[c]{@{}l@{}}178, 179, 182, 185, 186, \\ 193, 194, 213\end{tabular} & \begin{tabular}[c]{@{}l@{}}184--186, 188, 190, \\ 192--194, 222, 224, 230\end{tabular} \\
\bottomrule
\end{tabular}
\label{3D_non-symmorphic}
\end{table}

\subsection{Examples of 3D Nonsymmorphic Altermagnetism}\label{3D_nonsym}

The hourglass mechanism developed for the 2D wallpaper group $P4'gm'$ carries over directly to three dimensions. The role of the global symmetry $S$ is unchanged: composing $S$ with a nonsymmorphic spatial operation $\tilde g$ produces an antiunitary composite $\tilde g S$. The same-spin twofold degeneracies then live on the momentum manifold left invariant by $\tilde g S$, wherever $(\tilde g S)^2 = -1$. Such degeneracies are the analogs of Kramers-theorem-protected nodal lines{\cite{Dresselhaus1965Graphite,Young2015Dirac2D,Xie2021KramersNodalLine} and nodal planes \cite{Herring1937,Wu2018NodalSurface,Wilde2021NodalPlanes} in nonsymmorphic systems. The squared phase is set by the fractional translation in $\tilde g$: if $\tau \in \{x,y,z\}$ denotes the axis of the half-lattice translation, then $(\tilde g S)^2 = e^{i k_\tau}$, which equals $-1$ on the BZ-boundary plane $k_\tau = \pi$. The hourglass criterion is likewise unchanged: whenever this same-spin manifold and the opposite-spin manifold protected by the altermagnetic group's spin-flipping crystalline symmetries occupy different regions of the BZ, any momentum path linking them must host a symmetry-enforced band crossing. The two nonsymmorphic operations are then distinguished by the geometry of the $\tilde g S$-invariant manifold: for a glide reflection the invariant momentum lies along the mirror normal, so the same-spin manifold is a one-dimensional nodal line; for a $C_2$ screw rotation it fills the plane perpendicular to the rotation axis, so the same-spin manifold is a two-dimensional nodal plane. We illustrate both cases below, with one altermagnetic space group each. 

 As a glide-plane example, consider space group No.~100 (${P4bm}$), the 3D extension of the 2D wallpaper group No.~12 (${P4gm}$); its altermagnetic counterpart is $P4'bm'$, generated by the glide $\tilde M_x=\{m_x\mid 1/2,1/2,0\}$ (the partner glide $\tilde M_y$ follows from $\tilde M_x$ by the $C_4$ rotation and is therefore not independent). After shifting the origin to absorb the $\hat x$ component of the fractional translation, $\tilde M_x$ acts on momentum as
\begin{equation}
\tilde M_x : (k_x, k_y, k_z) \to (-k_x, k_y, k_z)\, e^{i k_y/2},
\end{equation}
so that composing with $S$ yields
\begin{equation}
\tilde M_x S : (k_x, k_y, k_z) \to (k_x, -k_y, -k_z)\, e^{i k_y/2}\mathcal{K}.
\end{equation}
Momentum invariance under $\tilde M_x S$ requires $k_y, k_z \in \{0, \pi\}$, where $k_x$ is free, picking out four lines along the mirror normal $\hat x$. Squaring gives $(\tilde M_x S)^2 = e^{i k_y}$, which equals $-1$ when $k_y = \pi$; the same-spin twofold degeneracy is therefore enforced along the two BZ-boundary lines $k_y = \pi$, $k_z \in \{0, \pi\}$, $k_x \in [-\pi, \pi]$ [magenta lines in Fig.~\ref{3D_nonsymmorphic}(a)]. The opposite-spin degeneracies sit on the BZ diagonal planes [green planes in Fig.~\ref{3D_nonsymmorphic}(a)]. A momentum path connecting a magenta line to a green plane therefore hosts an hourglass dispersion, generalizing the 2D mechanism in Fig.~\ref{nonsymmorphic}(a).

 As a screw-axis example, consider space group No.~94 (${P4_22_12}$), whose altermagnetic counterpart ${P4_2'2_12'}$ is generated by the screw axis $\tilde C_{2x} = \{C_{2x} \mid 1/2,1/2,1/2\}$ (the partner screw $\tilde C_{2y}$ follows by $C_4$ rotation). After an analogous origin shift that absorbs the perpendicular ($\hat y, \hat z$) components of the fractional translation, $\tilde C_{2x}$ acts on momentum as
\begin{equation}
\tilde C_{2x} : (k_x, k_y, k_z) \to (k_x, -k_y, -k_z)\, e^{i k_x/2},
\end{equation}
and composing with $S$ gives
\begin{equation}
\tilde C_{2x} S : (k_x, k_y, k_z) \to (-k_x, k_y, k_z)\, e^{i k_x/2}\mathcal{K}.
\end{equation}
Momentum invariance now requires $k_x \in \{0, \pi\}$, where $k_y$ and $k_z$ are free, two planes perpendicular to the rotation axis $\hat x$. Squaring gives $(\tilde C_{2x} S)^2 = e^{i k_x}$, which equals $-1$ when $k_x = \pi$; the same-spin twofold degeneracy is therefore enforced on the BZ-boundary plane $k_x = \pi$, and by the same $C_4$ relation the partner $\tilde C_{2y} S$ enforces it on $k_y = \pi$ [magenta planes in Fig.~\ref{3D_nonsymmorphic}(b)]. The opposite-spin degeneracies again sit on the BZ diagonal planes [green planes in Fig.~\ref{3D_nonsymmorphic}(b)], so any path linking a magenta plane to a green plane hosts an hourglass dispersion.

 The two examples above are representative of a much broader class. Table~\ref{3D_non-symmorphic} catalogs the 3D altermagnetic space groups that host either a glide reflection $\tilde M$ or a screw rotation $\tilde C_2$, organized by crystal system. As anticipated by the two examples, glide reflections enforce same-spin nodal lines while screw $\tilde C_2$ rotations enforce same-spin nodal planes.

 Glide reflections in orthorhombic crystals are an important exception, for which no symmetry-enforced hourglass arises. The reason follows from the criterion itself: an hourglass requires the same-spin and opposite-spin manifolds to occupy different regions of the BZ, since only then does a path linking them force a band crossing. For a screw rotation the two families lie on different BZ surfaces, as in ${P4_2'2_12'}$ [Fig.~\ref{3D_nonsymmorphic}(b)].
 For a glide reflection the situation is more delicate, because the same-spin manifold is only a one-dimensional nodal line. Such a line need not lie within an opposite-spin nodal plane, but in orthorhombic altermagnets it does: the opposite-spin nodal planes are fixed at $k_x=0,\pm\pi$ and $k_y=0,\pm\pi$, and the same-spin nodal lines protected by $\tilde{M}_z S$ run within these very planes. The same-spin and opposite-spin conditions are then met along the same line, promoting the twofold degeneracy to a fourfold one, so the hourglass is no longer guaranteed. Fig.~\ref{SG33} shows a candidate material that realizes this exceptional case.

 Accordingly, we exclude from the candidate list in Table~\ref{3D_non-symmorphic} any space group whose glide-enforced same-spin lines are embedded in its opposite-spin planes, such as the orthorhombic glide-plane groups above; this ensures that every listed space group satisfies the hourglass criterion. A detailed classification of altermagnetism in these 3D space groups is left to future work.

 Within Table~\ref{3D_non-symmorphic}, the opposite-spin nodal planes are fixed in the BZ for every space group except Nos.~76, 78, 85, and 86. These four form a special class, with point groups $\boldsymbol{C}_4$ (Nos.~76 and 78) and $\boldsymbol{C}_{4h}$ (Nos.~85 and 86) and halving subgroups $\boldsymbol{C}_2$ and $\boldsymbol{C}_{2h}$, respectively. An opposite-spin nodal plane is formed and pinned by a spin-flipping (effective) vertical mirror, but none of these groups' spin-flipping crystalline symmetries is such a mirror. As a result, their opposite-spin nodal planes are not fixed and may deform continuously in momentum space. Nevertheless, they are not entirely free. The fourfold rotation $C_{4z}$ is itself spin-flipping and enforces opposite-spin degeneracies along its invariant lines $(0,0,k_z)$ and $(\pm\pi,\pm\pi,k_z)$; the opposite-spin nodal surfaces may deform, but they must always pass through these $C_{4z}$-invariant lines. The same-spin degeneracies, in turn, arise from the nonsymmorphic symmetry of each group. For Nos.~76 and 78, the screw rotation $\tilde{C}_{2z}$ generates same-spin nodal planes on the BZ boundary $k_z=\pm\pi$. For Nos.~85 and 86, the glide reflection $\{M_z|1/2,1/2,0\}$ enforces same-spin degeneracies along the BZ-boundary lines $(\pm\pi,0,k_z)$ and $(0,\pm\pi,k_z)$. Because these same-spin nodal manifolds need not lie within the unfixed opposite-spin nodal planes, the two degeneracy families remain distinct. The hourglass criterion is therefore satisfied, and all four groups support symmetry-enforced hourglass band crossings.

\begin{figure}
\centerline{\includegraphics[width=0.65\textwidth]{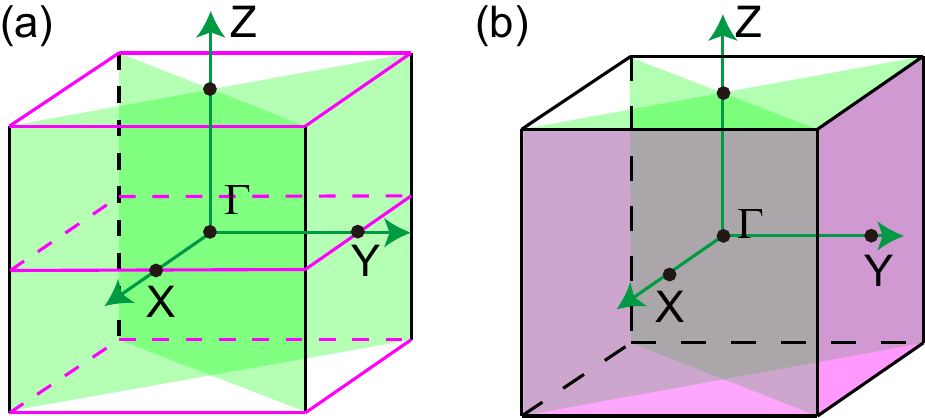}}
\caption{ (color online)
\textbf{Nonsymmorphic altermagnetism in the altermagnetic space groups P4$^\prime$bm$^\prime$ and $P4_2'2_12'$.}
 (a) and (b) BZs of altermagnetic space groups P4$^\prime$bm$^\prime$ and $ P4_2'2_12'$: Magenta lines and planes indicate same-spin twofold degeneracies, while the green planes represent opposite-spin twofold degeneracies.
}
\label{3D_nonsymmorphic}
\end{figure}

\section{Catalog of 3D altermagnetic space groups}

To exhaustively identify all possible altermagnetic orders within the 230 3D space groups, we begin by analyzing the 32 crystallographic point groups to determine which can serve as APGs. Our analysis identifies 37 distinct 3D APGs, comprising 10 centrosymmetric and 27 non-centrosymmetric groups. By incorporating translational symmetries, these APGs give rise to 422 distinct ASGs, including 160 centrosymmetric and 262 non-centrosymmetric cases, in agreement with the magnetic-space-group tabulation of Ref.~\cite{Yuan2021PRM} and with the direct type-III counting of Sec.~\ref{typeIII_count}. Beyond the conventional $d$-, $g$-, and $i$-wave classifications of altermagnetism, we systematically detail the momentum-space spin distributions within the BZ for all identified ASGs.

Similar to the 2D analysis, we apply two primary symmetry conditions to identify altermagnetic candidates among the 32 3D point groups: compensated collinear magnetic order and the breaking of $PT$ symmetry.

\begin{itemize}
    \item \textbf{Compensated collinear magnetic order:} To guarantee zero net magnetization, the point group $\boldsymbol{R}$ must contain a halving subgroup $\boldsymbol{H}$, which possesses the same cardinality (size) as its complementary coset $\boldsymbol{R} \setminus \boldsymbol{H}$. Because the point groups $\boldsymbol{C}_{1}$, $\boldsymbol{C}_{3}$, and $\boldsymbol{T}$ completely lack any such halving subgroups, they are fundamentally incompatible with altermagnetic ordering.

    \item \textbf{Breaking $PT$ symmetry:} The combined operation of inversion $P$ and time-reversal $T$ enforces Kramers degeneracy throughout the entire Brillouin zone, yielding a conventional antiferromagnetic order rather than a spin-split altermagnetic order. While the centrosymmetric point groups $\boldsymbol{S}_{2}$, $\boldsymbol{S}_{6}$, and $\boldsymbol{T}_{h}$ possess unique halving subgroups ($\boldsymbol{C}_{1}$, $\boldsymbol{C}_{3}$, and $\boldsymbol{T}$, respectively), these subgroups do not contain the inversion operation $P$. Consequently, $P$ is relegated to the spin-flipping coset $\boldsymbol{R} \setminus \boldsymbol{H}$. This arrangement intrinsically preserves $PT$ symmetry, enforcing full spin degeneracy and thereby ruling out these groups as candidates for altermagnetism.
\end{itemize}

The remaining 26 crystallographic point groups satisfy these essential symmetry prerequisites and are therefore capable of supporting altermagnetic ordering, as listed in Table~\ref{3D Point groups}. From these parent groups, we identify 37 distinct APGs out of the 58 possible 3D type-III magnetic point groups. While most point groups yield a single APG, certain groups, such as $\boldsymbol{D}_{3h}$ and $\boldsymbol{D}_{4h}$, contain multiple halving subgroups, each giving rise to two or three distinct APGs.

 A natural question is whether the spinless time-reversal symmetry $S = U_\pi T$ introduced earlier in the manuscript can rescue any of the six excluded point groups $\boldsymbol{C}_{1}$, $\boldsymbol{C}_{3}$, $\boldsymbol{T}$, $\boldsymbol{S}_{2}$, $\boldsymbol{S}_{6}$, $\boldsymbol{T}_{h}$, since $S$ is an additional non-spatial symmetry not yet incorporated into the screening above. The answer is no, and the reason is that in momentum space $S$ acts identically to spatial inversion $P$---both send $\boldsymbol{k} \to -\boldsymbol{k}$. The six excluded groups fall into two cases, one for each of the two screening conditions. The first three---$\boldsymbol{C}_{1}$, $\boldsymbol{C}_{3}$, $\boldsymbol{T}$---lack a halving subgroup. Augmenting each with $S$ produces, in momentum space, the direct-product equivalences
\begin{equation}
\boldsymbol{C}_{1} \otimes \widetilde{1} \cong \boldsymbol{S}_{2}, \qquad
\boldsymbol{C}_{3} \otimes \widetilde{1} \cong \boldsymbol{S}_{6}, \qquad
\boldsymbol{T} \otimes \widetilde{1} \cong \boldsymbol{T}_{h},
\end{equation}
where $\widetilde{1} = \{E, S\}$ and the equivalence holds because $S$ supplies an effective inversion in momentum space. The enlarged groups on the right are precisely the three centrosymmetric groups already excluded by the $PT$-Kramers argument, so the augmentation merely converts the ``no halving subgroup'' obstruction into the ``$PT$ symmetric'' obstruction without removing it. The remaining three---$\boldsymbol{S}_{2}$, $\boldsymbol{S}_{6}$, $\boldsymbol{T}_{h}$---already contain $P$. For these groups $S$ imposes no new constraint in momentum space (it acts the way $P$ already does), so the $PT$ symmetry that enforces full-BZ Kramers degeneracy survives and the spin-degenerate antiferromagnetic order persists. In both cases the exclusion is intact, leaving the 26 parent point groups identified above as the complete set capable of supporting altermagnetism.

Upon incorporating the global symmetry $S$, the 27 non-centrosymmetric APGs (highlighted in red in Table~\ref{3D Point groups}) become symmetry-equivalent to their centrosymmetric counterparts in momentum space. This equivalence is formally expressed as:
\begin{equation}
\text{Non-centrosymmetric group} \otimes \widetilde{1} \cong \text{Centrosymmetric group},
\label{equivalence}
\end{equation}
where taking the direct product of the non-centrosymmetric group with $\widetilde{1}=\{E, S\}$ yields its centrosymmetric counterpart. We emphasize that $\cong$ denotes an equivalence of the momentum-space action rather than a group isomorphism---$S$ is antiunitary while $P$ is unitary---but the identical action $\boldsymbol{k}\to-\boldsymbol{k}$ is the only property the screening conditions rely on. As a direct consequence of Eq.~\ref{equivalence}, non-centrosymmetric APGs and by extension, non-centrosymmetric ASGs inherit the exact spin-splitting topologies of their centrosymmetric analogs. It is worth noting that for APGs that inherently possess inversion symmetry, the inclusion of $S$ is redundant and imposes no additional symmetry constraints.

In general, altermagnetic spin configurations are classified into six fundamental wave types (the 2D and 3D variants of $d$-, $g$-, and $i$-waves). While the 2D variants were discussed in the 2D catalog section, their 3D counterparts are distinguished by momentum-dependent spin splitting that extends along all three spatial directions. Near the $\Gamma$ point, the spin-splitting energy of the low-energy effective Hamiltonian for these 3D waves can be described by spherical harmonics $Y_\ell^m(\theta,\phi)$. Consistent with the harmonic degrees $\ell = 2, 4, 6$ established in the 2D definitions, the 3D forms are expressed as:
\begin{align}
    \Delta E_{d\text{-wave}}^{3D} \sim & \left[ k^2Y_2^1(\theta,\phi)  \right] \sigma_z,\label{3Dd-wave} \\
    \Delta E_{g\text{-wave}}^{3D} \sim & \left[ k^4Y_4^3(\theta,\phi)  \right] \sigma_z, \label{3Dg-wave} \\
    \Delta E_{i\text{-wave}}^{3D} \sim & \left[ k^6Y_6^3(\theta,\phi) \right] \sigma_z, \label{3Di-wave}
\end{align}
Table~\ref{3D Point groups} summarizes the corresponding altermagnetic wave types for all APGs. By incorporating translational symmetries, the 37 identified APGs naturally generate 422 type-III magnetic space groups capable of supporting altermagnetic order, collectively referred to as 422 ASGs. These ASGs include both 160 centrosymmetric and 262 non-centrosymmetric cases, as listed in the Supplemental Material.

\begin{figure}
\centerline{\includegraphics[width=0.7\textwidth]{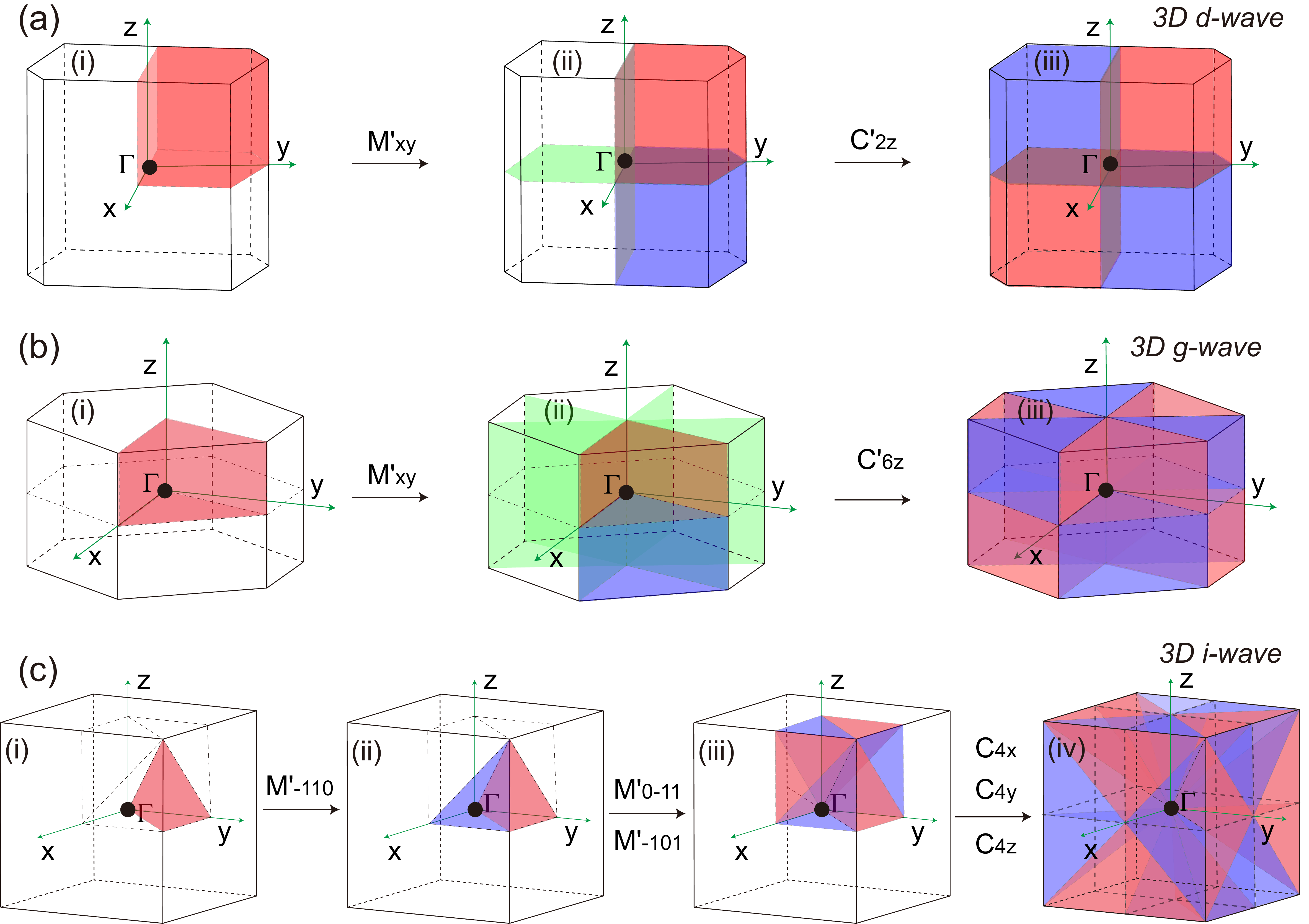}}
\caption{(color online)Generation process of the altermagnetic types in centrosymmetric ASGs (a) P2$^{\prime}$/m$^{\prime}$, (b) P6$^{\prime}$/m$^{\prime}$m$^{\prime}$m, (c) Pm$\bar{3}$m$^{\prime}$. Red and blue regions represent spin up and spin-down states in the BZs. Green and gray planes denote locked (symmetry-pinned) and unlocked spin-degenerate nodal planes, respectively. In (c), the six locked nodal planes are not highlighted in green, to keep the 3D spin distribution legible.
\label{3D_type} }
\end{figure}

In an altermagnet, the spin character of each band is locked to crystal momentum by the spin-group symmetries: at every $\boldsymbol{k}$ the sign of the spin splitting is fixed, and it alternates across the Brillouin zone in a pattern set by the magnetic order. The boundaries between the spin-up and spin-down regions are spin-degenerate nodal planes, of two kinds. A \emph{locked} nodal plane is pinned to a fixed high-symmetry plane by a spin-flipping reflection symmetry, so its position cannot move. An \emph{unlocked} nodal plane must be present due to the symmetry, but its location is unfixed; it only marks where the alternating spin pattern crosses zero, so its location is momentum dependent and can deform. We call an altermagnet \emph{spin-locking} when its Brillouin zone is partitioned entirely by locked nodal planes; if even a single unlocked plane is present, the altermagnet as a whole is \emph{non-spin-locking}. Among all 422 ASGs, 386 are spin-locking, whereas the remaining 36 are non-spin-locking.

To concretely illustrate these 3D wave types, we analyze three representative centrosymmetric ASGs as case studies: $P2^{\prime}/m^{\prime}$, a low-symmetry monoclinic group supporting a 3D $d$-wave altermagnetic order; $P6^{\prime}/m^{\prime}m^{\prime}m$, a high-symmetry hexagonal group exhibiting 3D $g$-wave altermagnetism; and $Pm\bar{3}m^{\prime}$, a cubic space group realizing a 3D $i$-wave order.

\begin{itemize}
    \item \textbf{Monoclinic $\boldsymbol{P2^{\prime}/m^{\prime}}$ (3D $\boldsymbol{d}$-wave):}
    Derived from the APG $2^{\prime}/m^{\prime}$, the halving subgroup for this low-symmetry group is $\boldsymbol{S}_2 = \{[E\|E], [E\|P]\}$, where $E$ is the identity operation and $P$ is spatial inversion. The corresponding coset contains the spin-flipping symmetries $C^{\prime}_{2z}=\left[U_{\boldsymbol{n}_0}(\pi) \| C_{2z} \right]$ and $M^{\prime}_{z}=\left[U_{\boldsymbol{n}_0}(\pi) \| M_{z} \right]$, which serve as the generators for the momentum-space spin texture. Beginning with an initial irreducible BZ hosting spin-up states (indicated in red) [Fig.~\ref{3D_type}(a-i)], the application of $M^{\prime}_{z}$ maps this region across the $xy$-plane to an adjacent irreducible wedge with spin-down polarization [Fig.~\ref{3D_type}(a-ii)]. Subsequently applying the $C^{\prime}_{2z}$ rotation extends this configuration across the entire BZ. The resulting spin distribution, depicted in Fig.~\ref{3D_type}(a-iii), exhibits a characteristic 3D $d$-wave symmetry. Notably, this pattern features two spin-degenerate nodal planes separating the red and blue volumes. One is locked, pinned to the horizontal $xy$-plane by $M^{\prime}_{z}$ [green plane in Fig.~\ref{3D_type}(a-ii)]; the other is unlocked, sitting at a momentum-dependent vertical location that is free to deform [gray plane in Fig.~\ref{3D_type}(a-ii)]. Because one unlocked plane is present, $P2^{\prime}/m^{\prime}$ is a non-spin-locking altermagnet.

    \item \textbf{Hexagonal $\boldsymbol{P6^{\prime}/m^{\prime}m^{\prime}m}$ (3D $\boldsymbol{g}$-wave):}
    Constructed from the APG $6^{\prime}/m^{\prime}m^{\prime}m$, this ASG possesses the halving subgroup $\boldsymbol{D}_{3d}$. Its corresponding coset includes the antiunitary symmetries $M^{\prime}_{z}$ and $C^{\prime}_{6z}$, as well as three vertical mirror symmetries ($M^{\prime}_{100}$, $M^{\prime}_{010}$, $M^{\prime}_{110}$) passing through the $K$ and $K^{\prime}$ points. Starting from a spin-up polarized irreducible BZ [Fig.~\ref{3D_type}(b-i)], the mirror $M^{\prime}_{z}$ maps it to a spin-down sector directly below the $xy$-plane [Fig.~\ref{3D_type}(b-ii)]. The sixfold rotation $C^{\prime}_{6z}$ then propagates these configurations to tile the full BZ. The emergent spin-splitting distribution, illustrated in Fig.~\ref{3D_type}(b-iii), displays the highly alternating 12-sector structure characteristic of a 3D $g$-wave altermagnet. Furthermore, the pattern features four spin-degenerate nodal planes [green planes in Fig.~\ref{3D_type}(b-ii)] that partition the spin-polarized volumes. All four are locked, pinned by $M^{\prime}_{z}$ and the three vertical mirror symmetries; since every nodal plane is locked, $P6^{\prime}/m^{\prime}m^{\prime}m$ is a spin-locking altermagnet.

    \item \textbf{Cubic $\boldsymbol{Pm\bar{3}m^{\prime}}$ (3D $\boldsymbol{i}$-wave):}
    This group is derived from the APG $m\bar{3}m^{\prime}$ and is characterized by the halving subgroup $\boldsymbol{T}_h$. The accompanying coset contains the fourfold rotational symmetries ($C_{4x}^{\prime}$, $C_{4y}^{\prime}$, $C_{4z}^{\prime}$) alongside the antiunitary mirror symmetries ($M^{\prime}_{\bar{1}10}$, $M^{\prime}_{0\bar{1}1}$, $M^{\prime}_{\bar{1}01}$). Selecting these coset elements as generators, we initiate the construction with a spin-up configuration in the irreducible BZ [Fig.~\ref{3D_type}(c-i)]. The application of $M^{\prime}_{\bar{1}10}$ maps this region to a spin-down alignment [Fig.~\ref{3D_type}(c-ii)]. Further application of $M^{\prime}_{0\bar{1}1}$ and $M^{\prime}_{\bar{1}01}$ extends this alternating pattern to fill a cubic subregion equivalent to one octant (1/8) of the full BZ [Fig.~\ref{3D_type}(c-iii)]. Finally, applying the three fourfold rotational symmetries propagates this subregion throughout the remaining BZ volume. As shown in Fig.~\ref{3D_type}(c-iv), the fully realized spin pattern exhibits a complex, symmetry-enforced 3D $i$-wave topology. The distribution is bounded by six locked nodal planes, which serve as the boundaries between regions of opposite-spin polarization. As all six are locked, $Pm\bar{3}m^{\prime}$ is a spin-locking altermagnet (the six planes are not drawn in green, to keep the 3D distribution legible).
\end{itemize}

\begin{table}
\centering
\renewcommand\arraystretch{1.6}
\setlength{\tabcolsep}{2.5mm}{
    \begin{tabular}{|*{6}{c|}}
        \hline
        Crystal systems & Point group $\boldsymbol{R}$  & SG  & Subgroup $\boldsymbol{H}$ & APG & Type of altermagnetism\\  \hline
         \multirow{2}{*}{Triclinic} & $\boldsymbol{C}{_{1}}$   & $\#1$ & $\times$ & $\times$ & $\times$
          \\\cline{2-6}
          & $\boldsymbol{S}{_{2}}$ & $\#2$ & $\times$ & $\times$ & $\times$
            \\\hline
          \multirow{3}{*}{Monoclinic} & {\clr $\boldsymbol{C}{_{2}}$}  &  {\clr $\#3-5$ } &  {\clr $\boldsymbol{C}{_{1}}$} & {\clr 2$^{\prime}$} & \multirow{3}{*}{\makecell[c]{ B-2 \\ (3D $d$-wave)}}
            \\\cline{2-5}
              &  {\clr $\boldsymbol{C}{_{1h}}$} &  {\clr $\#6-9$} &  {\clr $\boldsymbol{C}{_{1}}$} & {\clr m$^{\prime}$} &
              \\\cline{2-5}
              & $\boldsymbol{C}{_{2h}}$ & $\#10-15$ & $\boldsymbol{S}{_{2}}$ &  2$^{\prime}$/m$^{\prime}$ &
            \\\hline
           \multirow{3}{*}{Orthorhombic}  &  {\clr $\boldsymbol{D}{_{2}}$}  & {\clr $\#16-24$} &  {\clr $\boldsymbol{C}{_{2}}$} & {\clr 2$^{\prime}$2$^{\prime}$2}
           & \multirow{3}{*}{\makecell[c]{ P-2 \\ (2D $d$-wave)}}
           \\\cline{2-5}
              &   {\clr $\boldsymbol{C}{_{2v}}$ } & {\clr $\#25-46$} &  {\clr $\boldsymbol{C}{_{2}}$, $\boldsymbol{C}{_{1h}}$} & {\clr m$^{\prime}$m$^{\prime}$2, m$^{\prime}$m2$^{\prime}$ }&
              \\\cline{2-5}
              & $\boldsymbol{D}{_{2h}}$  &$\#47-74$ &  $\boldsymbol{C}{_{2h}}$  & m$^{\prime}$m$^{\prime}$m &
           \\\hline
          \multirow{7}{*}{Tetragonal} & {\clr $\boldsymbol{C}{_{4}}$}  &  {\clr $\#75-80$} &  {\clr $\boldsymbol{C}{_{2}}$} & {\clr 4$^{\prime}$ } & \multirow{3}{*}{\makecell[c]{ P-2 \\ (2D $d$-wave)}}
          \\\cline{2-5}
          & {\clr $\boldsymbol{S}{_{4}}$ }  & {\clr $\#81-82$} &  {\clr $\boldsymbol{C}{_{2}}$} & {\clr -4$^{\prime}$} &
               \\\cline{2-5}
          & $\boldsymbol{C}{_{4h}}$  &  $\#83-88$ &   $\boldsymbol{C}{_{2h}}$ & 4$^{\prime}$/m &
               \\\cline{2-6}
          & {\clr$\boldsymbol{D}{_{4}}$} & {\clr $\#89-98$} &  {\clr $\boldsymbol{D}{_{2}}$, $\boldsymbol{C}{_{4}}$} & {\clr 4$^{\prime}$22$^{\prime}$, 42$^{\prime}$2$^{\prime}$ }& \multirow{4}{*}{\makecell[c]{ P-2~~(2D $d$-wave)\\ P-4~~(2D $g$-wave)}}
               \\\cline{2-5}
           & {\clr$\boldsymbol{C}{_{4v}}$ }  & {\clr $\#99-110$} &  {\clr $\boldsymbol{C}{_{2v}}$, $\boldsymbol{C}{_{4}}$} & {\clr 4$^{\prime}$m$^{\prime}$m, 4m$^{\prime}$m$^{\prime}$ }&
               \\\cline{2-5}
          & {\clr$\boldsymbol{D}{_{2d}}$ }& {\clr $\#111-122$} &  {\clr $\boldsymbol{D}{_{2}}$, $\boldsymbol{C}_{2v}$, $\boldsymbol{S}{_{4}}$}& {\clr  -4$^{\prime}$2$^{\prime}$m, -4$^{\prime}$2m$^{\prime}$, -42$^{\prime}$m$^{\prime}$ } &
               \\\cline{2-5}
          & $\boldsymbol{D}{_{4h}}$ &   $\#123-142$ &   $\boldsymbol{D}{_{2h}}$, $\boldsymbol{C}{_{4h}}$ & 4$^{\prime}$/mm$^{\prime}$m, 4/mm$^{\prime}$m$^{\prime}$  &
           \\\hline
           \multirow{5}{*}{Trigonal} & $\boldsymbol{C}{_{3}}$   &   $\#143-146$ &  $\times$ &  $\times$ &  $\times$
             \\\cline{2-6}
           & $\boldsymbol{S}{_{6}}$  &$\#147-148$ &  $\times$ &  $\times$ &  $\times$
               \\\cline{2-6}
            & {\clr$\boldsymbol{D}{_{3}}$}  & {\clr $\#149-155$} &  {\clr $\boldsymbol{C}{_{3}}$} & {\clr 32$^{\prime}$ }& \multirow{3}{*}{\makecell[c]{ B-4 \\ (3D $g$-wave)}}
               \\\cline{2-5}
             & {\clr$\boldsymbol{C}{_{3v}}$}  & {\clr $\#156-161$}  &  {\clr $\boldsymbol{C}{_{3}}$} &{\clr 3m$^{\prime}$ }&
               \\\cline{2-5}
          & $\boldsymbol{D}{_{3d}}$ &  $\#162-167$ &  $\boldsymbol{S}{_{6}}$ & -3m$^{\prime}$ &
               \\\hline
          \multirow{7}{*}{Hexagonal} & {\clr $\boldsymbol{C}{_{6}}$ }&  {\clr $\#168-173$} &  {\clr $\boldsymbol{C}{_{3}}$} & {\clr 6$^{\prime}$} & \multirow{3}{*}{\makecell[c]{ B-4 \\ (3D $g$-wave)}}
             \\\cline{2-5}
           &  {\clr $\boldsymbol{C}{_{3h}}$} &  {\clr $\#174$} &  {\clr $\boldsymbol{C}{_{3}}$} & {\clr -6$^{\prime}$ }&
               \\\cline{2-5}
            & $\boldsymbol{C}{_{6h}}$  &   $\#175-176$ &  $\boldsymbol{S}{_{6}}$ & 6$^{\prime}$/m$^{\prime}$ &
               \\\cline{2-6}
            & {\clr$\boldsymbol{D}{_{6}}$ } &  {\clr $\#177-182$} &  {\clr $\boldsymbol{C}{_{6}}$, $\boldsymbol{D}{_{3}}$} & {\clr 62$^{\prime}$2$^{\prime}$, 6$^{\prime}$22$^{\prime}$} & \multirow{4}{*}{\makecell[c]{ P-6~~(2D $i$-wave)\\ B-4~~(3D $g$-wave)}}
               \\\cline{2-5}
            & {\clr$\boldsymbol{C}{_{6v}}$ } & {\clr $\#183-186$} &  {\clr $\boldsymbol{C}{_{6}}$, $\boldsymbol{C}{_{3v}}$}  & {\clr 6m$^{\prime}$m$^{\prime}$, 6$^{\prime}$mm$^{\prime}$} &
               \\\cline{2-5}
          &{\clr $\boldsymbol{D}{_{3h}}$} & {\clr $\#187-190$} &  {\clr $\boldsymbol{C}{_{3h}}$, $\boldsymbol{D}{_{3}}$, $\boldsymbol{C}{_{3v}}$}  & {\clr -6$^{\prime}$m2$^{\prime}$, -6$^{\prime}$m$^{\prime}$2, -6m$^{\prime}$2$^{\prime}$} &
                \\\cline{2-5}
          & $\boldsymbol{D}{_{6h}}$ & $\#191-194$ &  $\boldsymbol{C}{_{6h}}$, $\boldsymbol{D}{_{3d}}$ & 6/mm$^{\prime}$m$^{\prime}$, 6$^{\prime}$/m$^{\prime}$mm$^{\prime}$ &
              \\\hline
           \multirow{5}{*}{Cubic} &$\boldsymbol{T}$  &  $\#195-199$ &   $\times$ &   $\times$ &   $\times$
                \\\cline{2-6}
           & $\boldsymbol{T}{_{h}}$ & $\#200-206$ &   $\times$ &   $\times$ &   $\times$
               \\\cline{2-6}
             &{\clr$\boldsymbol{O}$ } & {\clr $\#207-214$} &  {\clr $\boldsymbol{T}$} & {\clr 4$^{\prime}$32$^{\prime}$} & \multirow{3}{*}{\makecell[c]{ B-6 \\ (3D $i$-wave)}}
                \\\cline{2-5}
           & {\clr$\boldsymbol{T}{_{d}}$ }& {\clr $\#215-220$ }    &  {\clr $\boldsymbol{T}$} & {\clr -4$^{\prime}$3m$^{\prime}$ } &
                       \\\cline{2-5}
           & $\boldsymbol{O}{_{h}}$ & $\#221-230$ &   $\boldsymbol{T}_h$ & m-3m$^{\prime}$ &
                \\\hline
    \end{tabular}}
    \caption{Classification of the corresponding altermagnetic point groups from the 32 3D crystallographic point groups. Point groups that are fundamentally incompatible with altermagnetic ordering are marked with `$\times$', whereas non-centrosymmetric groups and their associated halving subgroups/APGs are highlighted in red. The APGs, derived via Eq.~\ref{definition}, are expressed using standard magnetic point group notation. The parent crystallographic point groups $\boldsymbol{R}$ and their halving subgroups $\boldsymbol{H}$ are labeled using Schoenflies notation, specifically adopting the improper rotation convention for the $\boldsymbol{S}_n$ groups. The table additionally catalogs the corresponding space group (SG) indices and the altermagnetic wave topologies (e.g., 2D/3D $d$-, $g$-, and $i$-waves). The wave-type codes P-$n$ and B-$n$ denote planar (2D) and bulk (3D) spin distributions, respectively.}
    \label{3D Point groups}
\end{table}


\begin{table*}
\centerline{\includegraphics[width=0.85\textwidth]{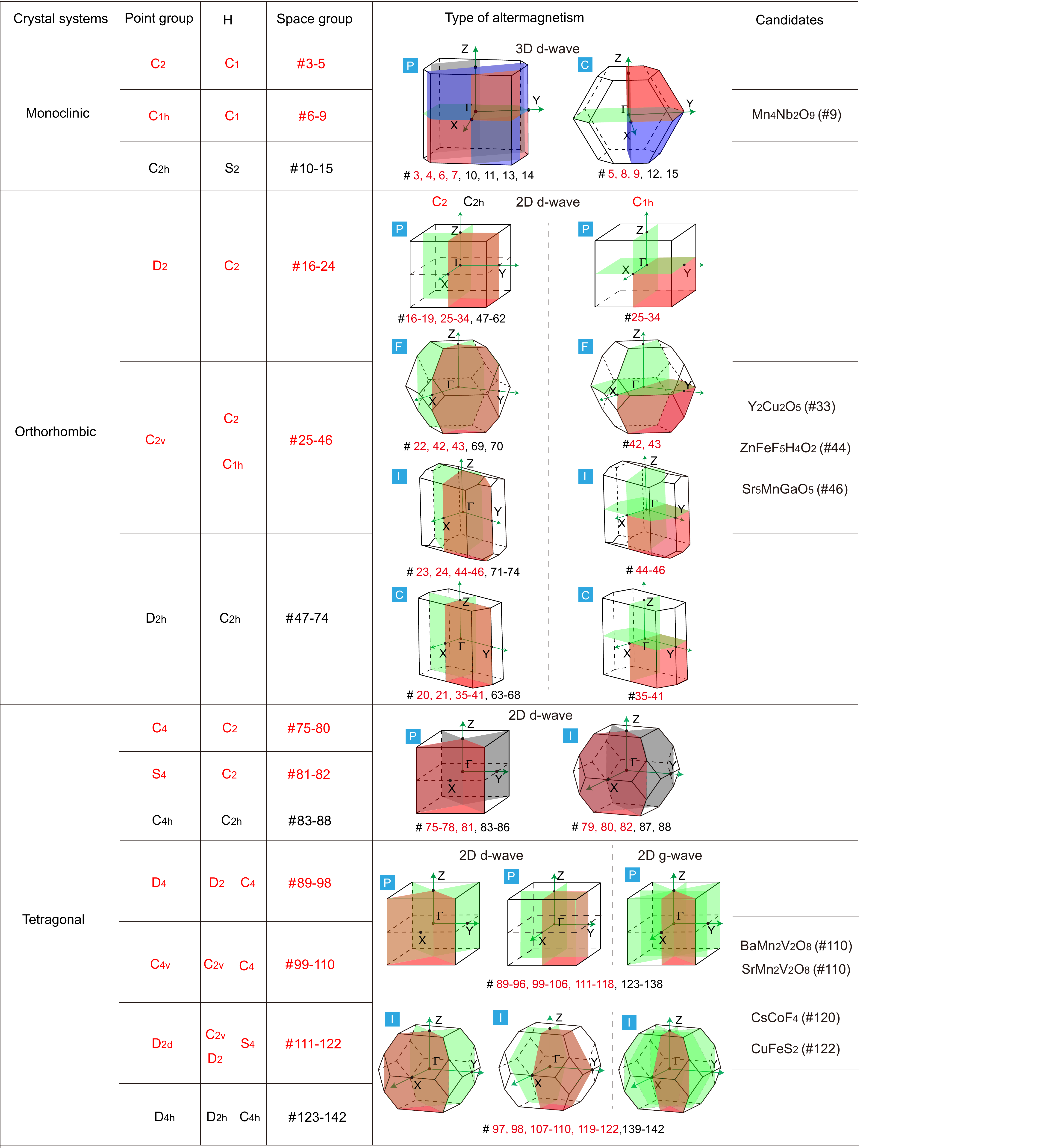}}
\caption{(color online)  Classification of altermagnetic types and the corresponding spin distributions in the BZ for space groups \#3--142 (monoclinic, orthorhombic, and tetragonal systems), together with the associated non-centrosymmetric candidate materials. The symbols ``P", ``C", ``I", ``F" and ``R" denote primitive, base-centered, body-centered, face-centered and rhombohedral Bravais lattices, and their corresponding BZs are presented using a unified coordinate system $(X,Y,Z)$. The green and gray planes represent locked and unlocked spin-degenerate nodal planes, respectively: green (locked) planes are spin-flipping mirror-symmetry-pinned and fixed in momentum space, while gray (unlocked) planes can deform. Nodal planes lying on the BZ-boundary faces are not highlighted in the drawings (see main text). The space groups corresponding to ASGs are listed in SM. It should be noted that the shapes of the BZ are not unique for certain Bravais lattices, including Monoclinic C, Orthorhombic F, Tetragonal I, and trigonal R. Only one representative Brillouin-zone construction is shown here, while other equivalent BZ shapes are provided in the SM.
\label{3D_1} }
\end{table*}

\begin{table*}
\centerline{\includegraphics[width=0.95\textwidth]{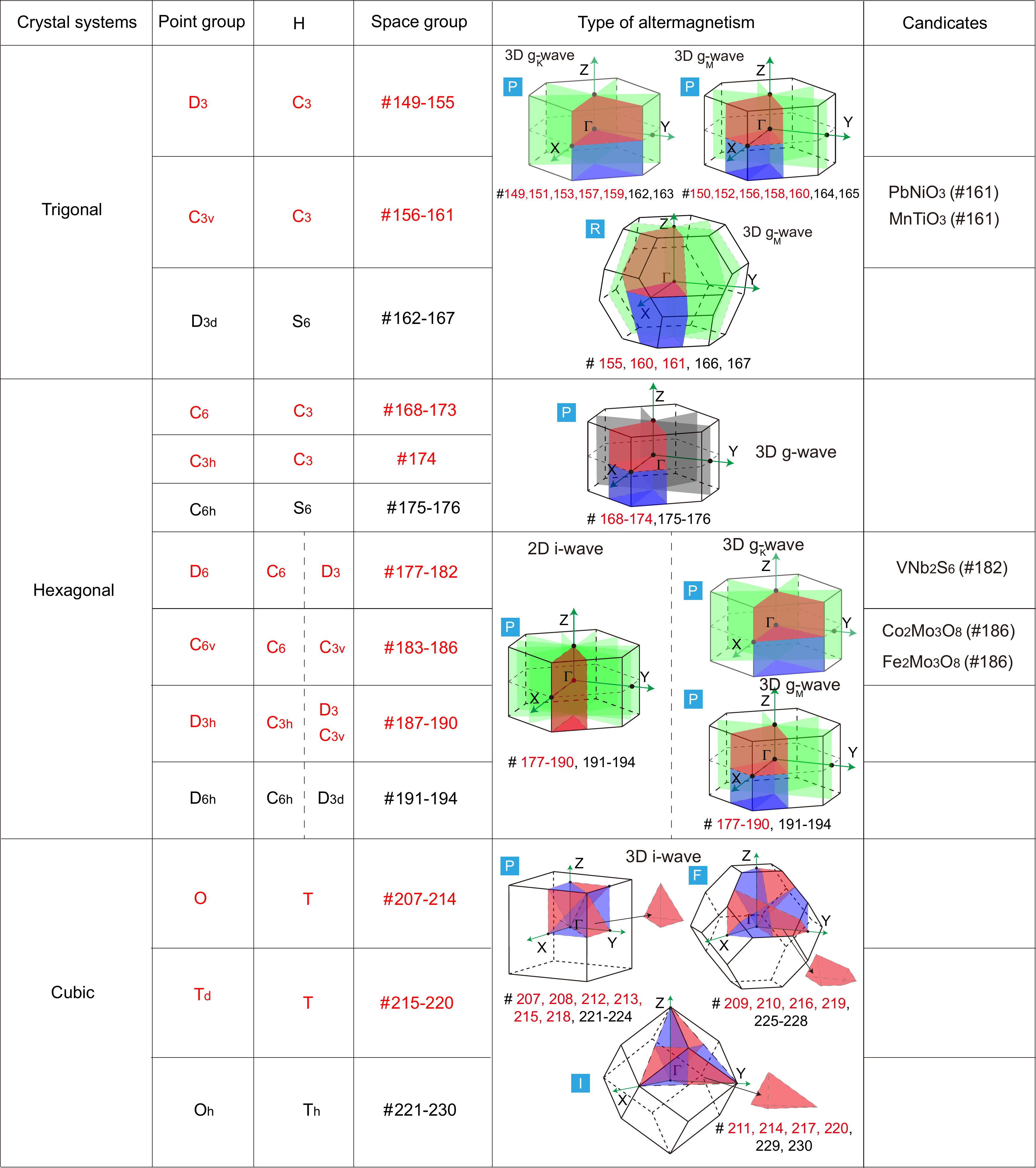}}
\caption{(color online) Continuation of Table~\ref{3D_1} for altermagnetic space groups \#149--230 (trigonal, hexagonal, and cubic systems), together with the associated non-centrosymmetric candidate materials. All symbols and color conventions follow Table~\ref{3D_1}.
\label{3D_2} }
\end{table*}

 Based on the above analysis, Table~\ref{3D_1} and Table~\ref{3D_2} present a comprehensive classification of all ASGs together with their characteristic spin distributions in the BZ, ordered by space group: Table~\ref{3D_1} covers space groups \#3--142 and Table~\ref{3D_2} covers \#149--230, together spanning all space groups capable of hosting altermagnetism. Each ASG is labeled by its corresponding conventional space group in the fourth column, and non-centrosymmetric altermagnetic groups are highlighted in red; the full correspondence between ASGs and type-III magnetic space groups is given in the Supplemental Material. For each crystal system, the BZs are constructed from the primitive cell in all cases and labeled by their Bravais lattice type. All BZs are drawn in a unified momentum-space coordinate system so that symmetry features can be compared directly across lattice types.

A given Bravais lattice can take several BZ shapes depending on its real-space geometry, so for each lattice type we display only one representative BZ in Table~\ref{3D_1} and Table~\ref{3D_2}, with the remaining geometries provided in the Supplemental Material. For example, the base-centered monoclinic (mC) lattice is partitioned into five BZ types---designated MCLC1 through MCLC5 in the computational literature~\cite{Setyawan2010}---based entirely on the relative ratios of the real-space lattice parameters ($a$, $b$, $c$) and the monoclinic angle $\beta$; we adopt MCLC3 as the representative here. The face-centered orthorhombic (ORCF2), body-centered tetragonal (BCT2), and rhombohedral (RHL1) lattices are each represented by a single BZ in the same manner.

 In the BZ diagrams, red and blue regions denote spin-up and spin-down states, and the interfaces between them are spin-degenerate nodal planes, shown as green (locked, symmetry-pinned) and gray (unlocked) planes. An altermagnet whose BZ is bounded entirely by green planes is spin-locking; the presence of even a single gray plane makes it non-spin-locking. In the cubic system every nodal plane is locked, so the system is fully spin-locking; there we omit the green highlighting to maintain visual clarity. Finally, the last column of the table lists candidate materials for the non-centrosymmetric altermagnetic groups, with the corresponding density functional theory calculation details provided in the Supplemental Material.

 The spin distribution is periodic in the reciprocal lattice: the spin states at $\boldsymbol{k}$ and $\boldsymbol{k}+\boldsymbol{G}$ coincide for any reciprocal-lattice vector $\boldsymbol{G}$. This periodicity adds two features to the nodal-plane structure that the drawings do not highlight. First, locked nodal planes recur on the BZ boundary: when a spin-flipping mirror pins a nodal plane through the BZ center, the parallel boundary faces of a primitive-lattice BZ are mapped onto themselves by the same mirror combined with a reciprocal-lattice translation, and are therefore pinned as well; to keep the three-dimensional drawings legible, these boundary nodal planes are not colored in Table~\ref{3D_1} and Table~\ref{3D_2}. Second, in non-primitive lattices the mirror-pinned planes alone are not always compatible with this periodicity, and additional spin-degenerate nodal surfaces are then required. In the body-centered orthorhombic BZ, for example, the shortest reciprocal-lattice vectors mix the three axes, so a spin pattern bounded only by the two locked planes $k_x=0$ and $k_y=0$ cannot be periodic, and additional unlocked nodal surfaces are enforced (minimally, a pair transverse to the $k_z$ axis). In the body-centered tetragonal BZ, by contrast, the square boundary faces are pinned by the diagonal spin-flipping mirrors and act as locked nodal planes, which restores the consistency without any additional unlocked surface. These periodicity-enforced degeneracies go beyond the point-group analysis of this section.


Because the non-spatial symmetry $S$ maps each non-centrosymmetric space group onto a symmetry-equivalent centrosymmetric one [Eq.~\ref{equivalence}], the two share the same altermagnetic spin distribution. Each BZ in Table~\ref{3D_1} and Table~\ref{3D_2} is therefore labeled by both: the centrosymmetric space group together with its red-marked non-centrosymmetric equivalents. For example, the non-centrosymmetric space groups \#25--34 share the spin distribution BZ of the centrosymmetric groups \#47--62.

In Table~\ref{3D_1} and Table~\ref{3D_2}, there exists a class of opposite-spin nodal planes that are not strictly fixed by symmetry. The corresponding point groups include all monoclinic point groups, the tetragonal point groups $\boldsymbol{C}_4$, $\boldsymbol{S}_4$, and $\boldsymbol{C}_{4h}$, as well as the hexagonal point groups $\boldsymbol{C}_6$, $\boldsymbol{C}_{3h}$, and $\boldsymbol{C}_{6h}$. A common feature of these point groups is that their cosets $\boldsymbol{R} \setminus \boldsymbol{H}$ lack vertical mirror symmetries, and therefore the opposite-spin nodal planes cannot be pinned to specific high-symmetry planes in momentum space. Nevertheless, the cosets $\boldsymbol{R} \setminus \boldsymbol{H}$ still contain rotational symmetries, such as $C_{2z}$, $C_{3z}$, or $C_{4z}$, which guarantee the existence of opposite-spin degeneracies along the corresponding rotation-invariant momentum lines, such as, along the $\Gamma$--Z line. As a consequence, although the exact shape and spatial position of these opposite-spin nodal planes are not unique, they remain constrained by the rotation-protected degeneracy lines and must pass through these invariant momentum lines. A detailed analysis of the monoclinic case is provided in the SM.

The following features summarize key aspects of the altermagnetic classification illustrated in Table~\ref{3D_1} and Table~\ref{3D_2}:

\begin{itemize}
\item{\textbf{BZ-geometry-dependent spin distributions}:  For a given point group and halving subgroup, all resulting ASGs share one altermagnetic point group, and hence one altermagnetic type. Nevertheless, these ASGs can exhibit different spin-distribution BZs, either because they belong to different Bravais lattices or because, even within a Bravais lattice, different lattice parameters give different BZ geometries. For instance, space groups \#47--74 all derive from the point group $\boldsymbol{D}_{2h}$ with halving subgroup  $\boldsymbol{C}_{2h}$, yielding the APG m$^{\prime}$m$^{\prime}$m, whose two spin-flipping (primed) mirrors pin locked (green, symmetry-protected) nodal planes at the $xz$ and $yz$ planes and produce the same spin-locking two-dimensional $d$-wave order throughout. Across the four orthorhombic Bravais lattices they form the P (Nos.~47--62), C (Nos.~63--68), F (Nos.~69--70), and I (Nos.~71--74) BZs---the P a regular cuboid, the others polyhedra. Even for the same Bravais lattice, moreover, the BZ geometry---and hence the appearance of the spin-distribution BZ---can vary with the real-space lattice parameters, as discussed above for the representative Brillouin zones (e.g.\ the five MCLC shapes of the base-centered monoclinic lattice).}

\item{\textbf{Role of halving subgroups:}
In the classification of altermagnetism, halving subgroups play two important roles: (1) certain point groups admit multiple halving subgroups, and hence the altermagnetic type is determined not only by the point group itself, but also by the chosen halving subgroup. (2) even for a fixed halving subgroup, different choices of symmetry generators can change the orientation of the nodal planes without changing the altermagnetic type, leading to different geometric patterns of the spin distribution. Thus, the number and location of mirror symmetries contained in the coset $\boldsymbol{R} \setminus \boldsymbol{H}$ associated with a given halving subgroup determine the number and orientation of opposite-spin degenerate manifolds, and determine the altermagnetic type. These two aspects are particularly evident in tetragonal and hexagonal crystal systems.

For the tetragonal point group \(\boldsymbol{D}_{4h}\), the cosets associated with the halving subgroups \(\boldsymbol{D}_{2h}\) and \(\boldsymbol{C}_{4h}\) generate two and four inter-spin nodal planes, corresponding to \(d\)-wave and \(g\)-wave altermagnetism, respectively. Furthermore, for halving subgroup \(\boldsymbol{D}_{2h}\), the two sets of vertical mirror symmetries in the coset may be chosen as either the \(xz\) and \(yz\) planes or two diagonal mirror planes. Although both choices belong to the same \(d\)-wave altermagnetic order, they lead to different spin-distribution patterns in the BZ.

Similarly, for the hexagonal point group \(\boldsymbol{D}_{6h}\), the cosets associated with the halving subgroups \(\boldsymbol{D}_{3d}\) and \(\boldsymbol{C}_{6h}\) generate three and six inter-spin nodal planes, corresponding to \(g\)-wave and \(i\)-wave altermagnetism, respectively. For the halving subgroup \(\boldsymbol{D}_{3d}\), the three sets of vertical mirror symmetries in the coset may also be chosen with different orientations: they can pass either through the (K)-point directions or through the (M)-point directions, resulting in distinct spin-distribution patterns in the BZ.

In the classification tables, vertical dashed lines in the third column distinguish different halving subgroups of the same point group, and the BZ spin distributions in the fourth column follow the same order. To further illustrate the possible spin-distribution patterns within the same altermagnetic type, the corresponding equivalent spin-distribution patterns are also shown in the fourth column.
\item{\textbf{Freedom to choose the principal axis in the halving subgroup:}}
In the orthorhombic crystal system, the principal axis in the halving subgroup that defines the altermagnetic type is not unique; it may be taken along the \(x\), \(y\), or \(z\) direction. For the point groups \(\boldsymbol{D}_2\) and \(\boldsymbol{D}_{2h}\), the choice is straightforward: the halving subgroup ($\boldsymbol{C}_2, \boldsymbol{C}_{2h}$) can be aligned with any of the three axes, and these orientations are permutationally equivalent, related simply by an exchange of axes. We therefore take the \(z\) axis---so that the spin distribution is invariant under the \(C_{2z}\) rotation---and display this representative in Table~\ref{3D_1}. The point group \(\boldsymbol{C}_{2v}=\{E,C_{2z},M_x,M_y\}\) is different: its three halving subgroups are genuinely distinct---the rotation subgroup \(\boldsymbol{C}_2=\{E,C_{2z}\}\) and the two mirror subgroups \(\boldsymbol{C}_{1h}^{x}=\{E,M_x\}\) and \(\boldsymbol{C}_{1h}^{y}=\{E,M_y\}\)---and are not related to one another by an exchange of axes. Once the spinless time-reversal symmetry \(S\) is included, however, their extensions coincide with the three orientations of the \(\boldsymbol{C}_{2h}\) group [Eq.~(\ref{equivalence})]: \(\boldsymbol{C}_2\otimes\widetilde{1}=\{E,C_{2z},\bar{P},\bar{M}_z\}\cong\boldsymbol{C}_{2h}^{z}\), \(\boldsymbol{C}_{1h}^{x}\otimes\widetilde{1}=\{E,\bar{C}_{2x},\bar{P},M_x\}\cong\boldsymbol{C}_{2h}^{x}\), and \(\boldsymbol{C}_{1h}^{y}\otimes\widetilde{1}=\{E,\bar{C}_{2y},\bar{P},M_y\}\cong\boldsymbol{C}_{2h}^{y}\). The three halving subgroups therefore share a single altermagnetic type, differing only in whether the principal axis points along \(z\), \(x\), or \(y\). Table~\ref{3D_1} shows only the representative patterns, with the principal axis oriented along \(z\) and \(x\). For completeness, the spin distributions for all three orientations--\(x\), \(y\), and \(z\)--are provided in the SM.}

\end{itemize}

 Starting from the 32 crystallographic point groups, we have identified the altermagnetic point groups and their associated space groups, and organized them into the $d$-, $g$-, and $i$-wave families. Their momentum-space spin distributions are cataloged in Table~\ref{3D_1} and Table~\ref{3D_2}, and characterized by the locked or unlocked character of their nodal planes (distinguishing spin-locking from non-spin-locking altermagnets), by the geometry-dependence of the BZ across different Bravais lattices, and by the equivalence between non-centrosymmetric and centrosymmetric groups under the global symmetry $S$. This catalog offers a basic classification of the spin-distribution BZ at the level of point-group symmetry. Once nonsymmorphic symmetry---screw axes and glide planes---is introduced (Sec.~\ref{3D_nonsym}), the spin-distribution layout can become considerably richer, since the additional symmetry enforces extra degeneracies such as further nodal lines or planes and hourglass-type dispersions; a complete treatment of these cases lies beyond the scope of this manuscript.

\section{Candidate Materials}

\begin{figure}[t]
\centerline{\includegraphics[width=1.0\textwidth]{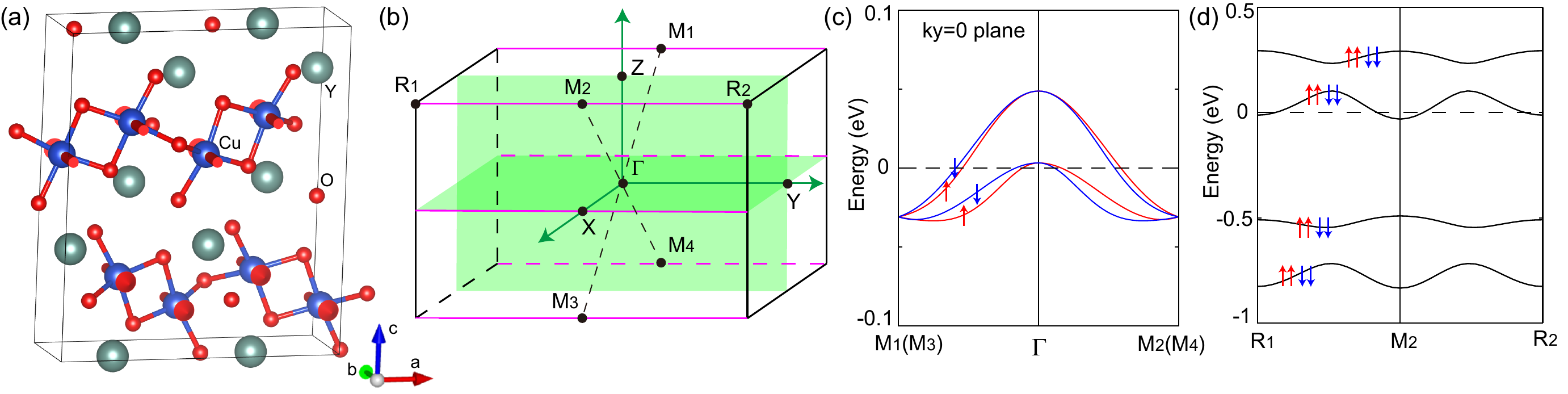}}
\caption{(color online) Non-centrosymmetric altermagnetic metal \ce{Y2Cu2O5}. (a) Crystal structure illustrating the magnetic spin arrangement. Red arrows denote the magnetic moments of the $\text{Cu}$ ions. (b) The corresponding bulk BZ. The green planes represent opposite-spin degeneracies, while magenta lines highlight high-symmetry paths where the bands exhibit fourfold degeneracy. (c) Calculated spin-resolved band structure along with high-symmetry lines. Red and blue lines represent spin up and down bands, respectively. (d) Calculated band structure along the high-symmetry line $R_1$-$M_2$-$R_2$. The black lines indicate fourfold degenerate bands, resulting from the superposition of same-spin and opposite-spin degeneracy mechanisms.
\label{SG33}}
\end{figure}

By using the MAGNDATA database~\cite{Gallego:ks5532,Gallego:ks5530} on the Bilbao Crystallographic Server, we find non-centrosymmetric candidate materials that can host altermagnetic order. These candidates are classified according to their point groups and corresponding space groups, as summarized in Table~\ref{3D_1} and Table~\ref{3D_2}. To illustrate the typical features of altermagnetism in non-centrosymmetric materials, we select \ce{Y2Cu2O5} compounds---identified as spin-split in a previous high-throughput MAGNDATA screening~\cite{Guo2023}---for detailed symmetry analysis and band structures. Symmetry analysis and band structures for additional candidate materials are provided in the Supplemental Material, providing a broader reference for the exploration of altermagnetic compounds.

As illustrated in Fig.~\ref{SG33}(a), the $\ce{Y2Cu2O5}$ compound\cite{Garca-Muoz1991} crystallizes in the nonsymmorphic space group $Pna2_1$ (No. 33). This space group features three key nonsymmorphic operations: $\tilde{C}_{2z}=\{C_{2z}|0~ 0~ 1/2\}$, $\tilde{M}_ y=\{M_y|1/2~ 1/2~ 0\}$, and $\tilde{M}_x=\{M_x|1/2~ 1/2~ 1/2\}$, where the tilde superscript is used to distinguish the nonsymmorphic operations from their corresponding point-group counterparts. Its associated point group $\boldsymbol{C}_{2v}$ contains a halving subgroup $\boldsymbol{C}_{1h}=\{E,~\tilde{M}_y\}$ that preserves the spin alignment within each sublattice. In contrast, the coset elements $\tilde{C}_{2z}$ and $\tilde{M}_x$ link sublattices with opposite spins, thereby enforcing opposite-spin degeneracy on the $yz$ planes. By combining $\tilde{C}_{2z}$ with the antiunitary symmetry $S$, an effective mirror symmetry $\bar{M}_z=\tilde{C}_{2z}S$ can be constructed. Since $\bar{M}_z$ connects sublattices with opposite spins, it protects the opposite-spin degenerate bands within the $xy$ plane. As highlighted as green planes in Fig.~\ref{SG33}(b), these opposite-spin degenerate $xy$ and $yz$ planes lead to a prominent fourfold symmetry distribution of spin splitting throughout the BZ. This characteristic identifies the material as a 2D $d$-wave altermagnet.

Furthermore, composing $\tilde{M}_y$ with $S$ yields the effective rotational symmetry $\bar{C}_{2y}=\tilde{M}_yS$, which connects sublattices with the same spin. This operator is found to satisfy $(\bar{C}_{2y})^2=(\tilde{M}_yS)^2=e^{ik_x}$. On the BZ boundary lines $(\pm\pi, k_y, 0)$ and $(\pm\pi, k_y, \pm\pi)$, the condition $(\tilde{M}_yS)^2=-1$ enforces Kramers-like same-spin twofold degeneracy. Since these paths simultaneously lie on the opposite-spin degenerate planes protected by $\tilde{M}_x$ and $\bar{M}_z$, the superposition of these two degeneracy mechanisms results in  fourfold degeneracy along these high-symmetry lines, marked as magenta lines in Fig.~\ref{SG33}(b). This fourfold degeneracy places \ce{Y2Cu2O5} in the exceptional class identified in Sec.~\ref{3D_nonsym}: in contrast to $P4'bm'$ and $P4_2'2_12'$, where the same-spin and opposite-spin degeneracy manifolds occupy different regions of the BZ and thereby enforce hourglass dispersions, here the same-spin nodal lines lie entirely within the opposite-spin nodal planes. The absence of this location discrepancy removes the hourglass criterion; instead, the two degeneracy mechanisms superpose, promoting the twofold degeneracies to fourfold ones.

To validate the aforementioned symmetry analysis, we calculated the non-relativistic band structure of $\text{Y}_2\text{Cu}_2\text{O}_5$ along the $M_1$-$\Gamma$-$M_2$ and $M_3$-$\Gamma$-$M_4$ paths in the $k_y=0$ plane, as illustrated in Fig.~\ref{SG33}(c). The calculations reveal that the spin-up and spin-down bands are mapped onto each other under the operations of $\bar{M}_{z}$ and $\tilde{M}_{x}$, which directly confirms the $d$-wave altermagnetism. Furthermore, Fig.~\ref{SG33}(d) displays the band structure along the high-symmetry line $R_1$-$M_2$-$R_2$. The results show that all bands along this path exhibit a distinct fourfold degeneracy. This observation strongly supports our theoretical analysis: the superposition of same-spin degeneracy (protected by $\tilde{M}_yS$) and opposite-spin degeneracy (protected by $\tilde{M}_{x}$) mechanisms jointly enforces the fourfold degenerate states along these high-symmetry lines.

\section{Conclusion and Outlook}

 We have established a complete catalog of altermagnetism for all magnetic wallpaper and space groups, based on the spin-splitting distributions in the BZ. The classification relies on two necessary symmetry conditions: the presence of a halving subgroup guarantees the compensated collinear magnetic order, and the absence of the combined symmetry of time reversal and inversion avoids Kramers degeneracy. These two conditions identify 17 altermagnetic wallpaper groups (12 centrosymmetric and 5 non-centrosymmetric) and 422 altermagnetic space groups (160 centrosymmetric and 262 non-centrosymmetric). Furthermore, beyond the point-group classification, the catalog presents the spin distribution in the primitive-cell BZ for each group; the spin-degenerate nodal lines and planes are either locked (pinned by spin-flipping mirror symmetries) or unlocked (free to deform). Among the 422 altermagnetic space groups, 386 are spin-locking and 36 are non-spin-locking.

The spinless time-reversal symmetry $S$ plays the key role throughout this classification. In momentum space, $S$ acts as an effective inversion and preserves identical spin states at opposite momenta even in the absence of crystal inversion; hence, altermagnetism extends to non-centrosymmetric crystals. In particular, each non-centrosymmetric altermagnetic group is symmetry-equivalent to a centrosymmetric counterpart and inherits its spin-splitting BZ. Furthermore, in nonsymmorphic crystals, the combination of $S$ and a glide mirror or a screw axis $\tilde{g}$ forms the antiunitary symmetry $\tilde{g}S$, which enforces same-spin Kramers degeneracy on the BZ boundary. The compatibility relations between this same-spin degeneracy and the opposite-spin degeneracy of the conventional altermagnetic order lead to symmetry-enforced hourglass dispersions. Therefore, the spin-splitting patterns extend beyond the six wave types ($d$-, $g$-, and $i$-wave in 2D and 3D) reported in the literature.

In two dimensions, only one magnetic wallpaper group hosts nonsymmorphic altermagnetism, where the hourglass dispersions lead to more complicated spin-alternating patterns in the BZ. However, three dimensions are much more fruitful. As cataloged in Table~\ref{3D_non-symmorphic}, numerous altermagnetic space groups in the orthorhombic, tetragonal, hexagonal, and cubic crystal systems host glide or screw symmetries satisfying the hourglass criterion; the glide reflections enforce same-spin nodal lines, while the screw rotations enforce same-spin nodal planes. We leave the detailed classification of the spin-distribution BZs in these 3D nonsymmorphic magnetic space groups for future study. Our catalog serves as a guide for the discovery of altermagnetic materials beyond centrosymmetric crystals and opens a door for exploring novel band structures rooted in nonsymmorphic symmetries.

\medskip
\textbf{Supporting Information} \par 
Supporting Information is available from the Wiley Online Library or from the author.

\medskip
\textbf{Acknowledgements} \par 
C.-K. C. was supported by Japan Science and Technology Agency (JST) as part of Adopting Sustainable Partnerships for Innovative Research Ecosystem, Grant No. JPMJAP2318, and by JST Presto Grant No. JPMJPR2357. X.W. is supported by the National Key R\&D Program of China (Grant No. 2023YFA1407300) and the National Natural Science Foundation of China (Grants No. 12574151, 12447103 and 12447101). C.C.L. is supported by the start-up fund from Hefei National Laboratory and the RIKEN TRIP initiative (RIKEN Quantum). 

\medskip

%
\bibliographystyle{MSP}
\bibliography{Alter_references}

\end{document}